\documentclass{aa}

\usepackage[varg]{txfonts}

\usepackage{graphicx}
\usepackage{tabularx}
\usepackage{float}

\usepackage{epsfig}
\usepackage{longtable}
\usepackage{supertabular}
\usepackage{color}
\usepackage{amssymb}
\usepackage{amsmath}
\usepackage{textcomp}
\usepackage{natbib}

\usepackage{hyperref} 

\begin{document} 

\newcommand{\ms}{m\,s$^{-1}$\/}
\newcommand{\kms}{km\,s$^{-1}$\/}
\newcommand{\cmss}{cm\,s$^{-2}$\/}
\newcommand{\gcm}{g\,cm$^{-3}$\/}

\title{The GAPS Programme with HARPS-N at TNG.
    \thanks{Based on observations collected at the Italian {\em Telescopio Nazionale Galileo} (TNG), 
    operated on the island of La Palma by the Fundaci\'on Galileo Galilei of the Istituto Nazionale di Astrofisica (INAF) at the Spanish 
    Observatorio del Roque de los Muchachos of the Instituto de Astrof\'isica de Canarias, in the frame of the programme 
    {\em Global Architecture of Planetary Systems} (GAPS).}}
 
\subtitle{XIII. The orbital obliquity of three close-in massive planets hosted by dwarf K-type stars: WASP-43, HAT-P-20 and Qatar-2.
    \thanks{Also based on observations collected at the  0.82m IAC80 Telescope,  operated on the island of Tenerife by the Instituto de
    Astrof\'isica de Canarias in the Spanish Observatorio del Teide.}}

\author{ M.~Esposito\inst{1}
          \and E.~Covino \inst{1} 
          \and S.~Desidera \inst{2}
          \and L.~Mancini \inst{16,3,6}
          \and V.~Nascimbeni \inst{10,2}
          \and R.~Zanmar Sanchez \inst{4}
          \and K.~Biazzo \inst{4}
          \and A.~F.~Lanza \inst{4}
          \and G.~Leto \inst{4}
          \and J.~Southworth \inst{5}
          \and A.~S.~Bonomo \inst{6}
          \and A.~Su\'{a}rez Mascare\~{n}o \inst{7,8}
          \and C.~Boccato \inst{2}
          \and R.~Cosentino \inst{11}
          \and R.~U.~Claudi \inst{2}
          \and R.~Gratton \inst{2}
          \and A.~Maggio \inst{9}
          \and G.~Micela \inst{9}
          \and E.~Molinari \inst{11,14}
          \and I.~Pagano \inst{4}
          \and G.~Piotto \inst{10,2}
          \and E.~Poretti \inst{12}
          \and R.~Smareglia \inst{13}
          \and A.~Sozzetti \inst{6}
          \and L.~Affer \inst{9}
          \and D.R.~Anderson \inst{5}
          \and G.~Andreuzzi \inst{11,15}
          \and S.~Benatti \inst{2}
          \and A.~Bignamini \inst{13}
          \and F.~Borsa \inst{12}       
          \and L.~Borsato \inst{10,2}
          \and S.~Ciceri \inst{3}
          \and M.~Damasso \inst{6}
          \and L.~ di Fabrizio \inst{11}
          \and P.~Giacobbe \inst{6}
          \and V.~Granata \inst{10,2}
          \and A.~Harutyunyan \inst{11}
          \and T.~Henning \inst{3}
          \and L.~Malavolta \inst{10,2}
          \and J.~Maldonado \inst{9}
          \and A.~Martinez Fiorenzano \inst{11}
          \and S.~Masiero \inst{2}
          \and P.~Molaro \inst{13}
          \and M.~Molinaro \inst{13}
          \and M.~Pedani \inst{11}
          \and M.~Rainer \inst{12}
          \and G.~Scandariato \inst{4}
          \and O.D.~Turner \inst{5}
          }
          
\institute{ INAF -- Osservatorio Astronomico di Capodimonte, via Moiariello, 16, 80131 Naples, Italy
       \and INAF -- Osservatorio Astronomico di Padova, Vicolo dell'Osservatorio 5,  35122 Padova, Italy
       \and Max-Planck-Institut f\"ur Astronomie, K\"onigstuhl 17, D-69117 Heidelberg, Germany
       \and INAF -- Osservatorio Astrofisico di Catania, via S. Sofia 78, 95123 Catania, Italy  
       \and Astrophysics Group, Keele University, Staffordshire, ST5 5BG, UK 
       \and INAF -- Osservatorio Astrofisico di Torino, Via Osservatorio 20, I-10025, Pino Torinese, Italy 
       \and Instituto de Astrof\'isica de Canarias, C/ V\'ia L\'actea, s/n, E38205 - La Laguna (Tenerife), Spain 
       \and Departamento de Astrof\'isica, Universidad de La Laguna, Avda. Astrof\'isico Francisco  S\'anchez, s/n, 38206 La Laguna (TF), Spain
       \and INAF -- Osservatorio Astronomico di Palermo, Piazza del Parlamento, 1, 90134 Palermo, Italy 
       \and Dipartimento di Fisica e Astronomia Galileo Galilei -- Universit\`a di Padova , Vicolo dell’Osservatorio 2, 35122 Padova, Italy 
       \and INAF -- Fundaci\'{o}n  Galileo  Galilei,  Rambla  Jos\'{e}  Ana  Fernandez P\'{e}rez 7, 38712 Bre\~{n}a Baja, Spain  
       \and INAF -- Osservatorio Astronomico di Brera, via E. Bianchi 46, 23807 Merate (LC), Italy 
       \and INAF -- Osservatorio Astronomico di Trieste, via G. B. Tiepolo 11, 34143 Trieste, Italy 
       \and INAF - IASF Milano, via Bassini 15, 20133, Milano, Italy 
       \and INAF - Osservatorio Astronomico di Roma, via Frascati 33, 00040 Monte Porzio Catone (Roma), Italy 
       \and Dipartimento di Fisica, Universit\`a di Roma Tor Vergata, Via della Ricerca Scientifica 1, 00133  Roma, Italy 
}

\date{Received date /
Accepted date }

\abstract {The orbital obliquity of planets with respect to the rotational axis of their host stars
 is a relevant parameter for the characterization of the global architecture of planetary systems and a key
observational constraint to discriminate between different scenarios proposed to explain the existence of close-in giant planets.} 
{In the framework of the GAPS project, we are conducting an observational programme aimed at the
determination of the orbital obliquity of known transiting exoplanets. The targets are selected
to probe the obliquity against a wide range of stellar and planetary physical parameters.} 
{We exploit high-precision radial velocity (RV) measurements, delivered by the 
HARPS-N spectrograph at the 3.6m Telescopio Nazionale Galileo, to measure the Rossiter-McLaughlin (RM)
effect in RV time-series bracketing planet transits, and to refine the orbital parameters determinations
with  out-of-transit RV data. We also analyse new transit light curves obtained with several 1-2m class telescopes
to better constrain the physical fundamental parameters of the planets and parent stars.} 
{ We report here on new transit spectroscopic observations for three 
very massive close-in giant planets: WASP-43\,b, HAT-P-20\,b and Qatar-2\,b ($M_{\rm p}$ = 2.00, 7.22, 2.62\,M$_{\rm J}$;
$a$ = 0.015, 0.036, 0.022\,AU, respectively) orbiting dwarf K-type stars with effective temperature
well below 5000\,K ($T_{\rm eff}$ = 4500$\pm$100, 4595$\pm$45, 4640$\pm$65\,K respectively).
These are the coolest stars (except for WASP-80) for which the RM effect has been observed so far.
We find  $\lambda$ = 3.5$\pm$6.8 deg for WASP-43\,b and $\lambda$ = -8.0$\pm$6.9 deg for HAT-P-20\,b, while for Qatar-2, our faintest target,
the RM effect is only marginally detected, though our best-fit value $\lambda$ = 15$\pm$20\,deg is in agreement with a previous
determination.  In combination with stellar rotational periods derived photometrically, we estimate the true spin-orbit angle, 
finding that WASP-43\,b is aligned while the orbit of HAT-P-20\,b presents a small but significant
obliquity ($\varPsi$=36 $_{-12}^{+10}$\,deg).
By analyzing the CaII\,H\&K chromospheric emission lines for HAT-P-20 and WASP-43,
we find evidence for an enhanced level of stellar activity which is possibly induced by star-planet interactions.
}
{}


\titlerunning{RM effect in WASP-43, HAT-P-20 and Qatar-2}
\authorrunning{M.~Esposito et al.}
\maketitle

\section{Introduction}\label{Sec:Intro}

The observational study of extrasolar planets flourished in the last
two decades and has led to many surprising discoveries that challenged the 
traditional paradigms of planetary systems formation and evolution.
Among those discoveries are the facts that,
at variance with what is seen in our Solar System,
giant planets can exist very close to their parent stars ($a$<0.1 AU)
and their orbital planes can have large
obliquity angles with respect to the host star equatorial planes.

Information on the obliquity of exoplanets is more easily accessed 
when they happen to transit the host star. Several techniques have been
exploited to estimate the orbital obliquity of transiting planets (TPs):
star spot crossing \citep{2011ApJ...743...61S, 2014MNRAS.443.2391M}, asteroseismology \citep{2013ApJ...766..101C},
and gravitational darkening \citep{2011ApJS..197...10B}.
However the great majority of the $\sim$100 TPs obliquity measurements\footnote{We refer 
to http://www.astro.keele.ac.uk/jkt/tepcat/rossiter.html
for an updated list of published papers on obliquity measurements.}
 were obtained by observing the Rossiter-McLaughlin (RM) effect
 \citep{1924ApJ....60...15R, 1924ApJ....60...22M}. 
The RM effect is an anomaly in the RV orbital trend occurring
during transit, when the planet, by blocking part of the stellar light, causes a distortion of the spectral lines.
The shape of the RM anomaly is related to the sky-projected angle $\lambda$
between the planet orbital axis and the star spin.
For a more detailed description of the RM effect see, e.g., 
\citet{2000A&A...359L..13Q, 2005ApJ...622.1118O, 2006ApJ...650..408G, 2010ApJ...709..458H, 2013A&A...550A..53B, 2015MNRAS.454.4379B}.

Based on about 50 RM measurements, \citet{2012ApJ...757...18A}
(see also \cite{2010ApJ...718L.145W} and \cite{2010ApJ...719..602S}) 
noticed an empirical $\lambda$-$T_{\rm eff}$ trend: stars with
$T_{\rm eff}\gtrsim$6250 K have a broad distribution of obliquities
while stars with $T_{\rm eff}\lesssim$6250 K are aligned.
Most recent RM measurements have substantially confirmed the trend,
even though the  transition from aligned to oblique orbits appears to be a
smoother function of $T_{\rm eff}$, and some remarkable
exceptions have been discovered \citep{2014A&A...564L..13E,2010ApJ...723L.223W}.

The NASA Kepler mission has represented a major breakthrough
in the field of extrasolar planets science \citep{2016RPPh...79c6901B}.
As of now, the number of Kepler candidate transiting planets
is larger than 4600 \citep{2016ApJS..224...12C}. Even though the majority of them
lack the precise RV follow-up needed
for the planet mass determination, they constitute
a valuable statistical sample and allow to infer 
important general properties of exoplanetary systems.
Kepler found that systems with many transiting planets are common 
\citep{2011ApJS..197....8L}. Simple geometrical considerations lead
to believe that the orbits of such planets must be nearly coplanar;
\citet{2014ApJ...790..146F}, from the analysis of their transit duration ratios,
infer that the statistical mode of the orbital mutual inclinations  is in the range 1.0–2.2 deg.
On the other hand, \citet{2014ApJ...796...47M}, combining the information on stellar rotational period and
projected velocity of 70 Kepler objects of interest (KOIs), found with 95\% confidence that 
the obliquities of stars with one
transiting planet are systematically larger than those with multiple transiting planets, therewith suggesting that
single planets represent dynamically hotter systems than
the flat multiple transiting systems. 
\citet{2015ApJ...801....3M} compared the observed amplitude 
of the rotational photometric modulation of 993 KOIs
with 33,614 single Kepler stars in the temperature range of 3500–6500 K.
They found the amplitudes to be statistically higher for KOIs with
3500 K <$T_{\rm eff}$< 6000 K and lower for 6000 K <$T_{\rm eff}$< 6500 K,
and interpret this as an indication that cool TP host stars are aligned
while hot stars tend to have high obliquities. Their result is in line with
what is found by means of RM measurements.

The attempts to explain the observed obliquity distribution of exoplanets
 have addressed many fundamental open questions
of planet formation and evolution, as well as of physics
of star interiors. Do giant planets migrate inward
by effect of tidal interaction within the protoplanetary disc \citep{1996Natur.380..606L,2014prpl.conf..667B}
or following planet-planet scattering,
Kozai-Lidov cycles, and secular chaotic orbital evolution 
\citep{2013ApJ...767L..24D,2011ApJ...735..109W}?
Are there mechanisms, such as chaotic star formation  \citep{2010MNRAS.401.1505B,2011MNRAS.417.1817T,2012Natur.491..418B} 
or stellar internal gravity waves \citep{2012ApJ...758L...6R}, able to
misalign the protoplanetary disc plane and the equatorial stellar plane 
with respect to each other? How effective are star-planet interactions at re-orientating
originally misaligned systems \citep{2014ARA&A..52..171O,2012MNRAS.423..486L}?
Theoretical efforts will benefit from the characterization of orbital obliquities
in correspondence of a wider range of the relevant parameters, such as orbital separation
and eccentricity, star and planet mass, stellar effective temperature and metallicity, etc.

We are conducting systematic observations of 
known transiting planets 
in the frame of the programme {\em Global Architecture of Planetary Systems} (GAPS)
\citep{2013A&A...554A..28C, 2013A&A...554A..29D, 2016MmSAI..87..141P}.
 The selection of our targets is based on the following criteria:
i) we give priority to hot-Jupiter systems that allow us to widen the explored range of stellar characteristics 
(i.e., effective temperature and $\log g$) and planet orbital and physical parameters;
ii) in order to guarantee a reliable measurement of the RM effect, we restrict our choice to stars with visual magnitude $V<14$\,mag.
We have already presented measurements of the RM effect for six targets
in several papers of the GAPS series: Qatar-1b \citep{2013A&A...554A..28C},
HAT-P-18b \citep{2014A&A...564L..13E}, XO-2b \citep{2015A&A...575A.111D},
KELT-6b \citep{2015A&A...581L...6D}, HAT-P-36b and WASP-11b \citep{2015A&A...579A.136M}.

Here we report new observations of the RM effect for three other  
TPs: WASP-43\,b \citep{2011A&A...535L...7H}, HAT-P-20\,b \citep{2011ApJ...742..116B} 
and Qatar-2\,b \citep{2012ApJ...750...84B}. 
All of them are very massive close-in giant planets
(WASP-43\,b:$M_{\rm p}$ = 2 M$_{\rm J}$, $a$ = 0.015 AU;
HAT-P-20\,b: 7.2 M$_{\rm J}$, 0.036 AU; Qatar-2\,b: 2.6 M$_{\rm J}$, 0.022 AU),
hosted by dwarf K-type stars with effective temperature
well below 5000 K ($T_{\rm eff}$ = 4500$\pm$100, 4595$\pm$45, 4640$\pm$65 K respectively).
In fact, with the exception of WASP-80 \citep{2015MNRAS.450.2279T},
these are the three coolest TP host stars for which
the RM effect has been successfully observed.
No additional planet is known to orbit around the three host stars,
however for HAT-P-20 a long-term RV linear trend was observed
\citep{2014ApJ...785..126K,2015ApJ...805..132D}, possibly caused
by a stellar visual companion. Lucky-Imaging observations \citep{2015A&A...579A.129W}
put constraints on possible wide companions of WASP-43 and Qatar-2.
The high mass of the planets and the small orbital separations also qualify
the three systems as interesting candidates to investigate possible
stellar activity enhancement induced by the planets.


\section{Observations and data reduction}\label{Sec:ObsRed}

\subsection{Spectroscopic data}
All the spectra used in this work were acquired with the HARPS-N spectrograph
(wavelength coverage: 383-690 nm, resolving power $R$=115\,000),
installed at the TNG telescope \citep{2012SPIE.8446E..1VC}. 
The data were reduced by means of the latest version of the HARPS-N Data
Reduction Software (DRS) \citep{2014SPIE.9147E..8CC,2014ASPC..485..435S}.
In addition to 1-D wavelength-calibrated spectra, the DRS provides
radial velocities (RVs), calculated by cross-correlating spectra with a numerical mask
\citep{1996A&AS..119..373B,2002A&A...388..632P,2007A&A...468.1115L},
and line bisectors. 
The DRS measures also the Mount Wilson S index and, if the stellar B$-$V colour index
is lower than 1.2, also the log($R'_{\rm HK}$) chromospheric activity index is derived \citep{2011arXiv1107.5325L}.

\paragraph{\bf{WASP-43.}}
We acquired a time series of 32 spectra, bracketing the transit of WASP-43\,b occurring
on the 2013 March 11-12 night. With an exposure time of $T_{exp}$=7.5 minutes, the 
spectra have a signal-to-noise ratio (S/N) (per pixel in 1-D spectra at 5500 \AA) ranging from 12 to 20.
Between March 2013 and May 2015 
we acquired additional, out-of-transit, spectra at 8 different epochs.
The RV measurements
were obtained using a K5 mask. 
A log of the transit observations is reported in Table\,\ref{t:logspec} and all the RVs are shown in  Table\,\ref{t:WASP43-RV-data}.

\paragraph{\bf{HAT-P-20.}}
A complete transit of HAT-P-20\,b was observed on 2014 March 11-12; 
the time series of 23 spectra, with  exposure times of $T_{exp}$=10 minutes, started $\sim$1 hour 
before ingress and ended $\sim$1 hour after egress. The S/N (per pixel in 1-D spectra at 5500 \AA)
degraded from $\sim$25 to $\sim$20 during the series, with the increase of the airmass.
During the observations the target was close (11\,deg) to the almost full (81\,\%) Moon,
and the Moon RV differed from the star RV by only $\sim$8 \kms, that is $\sim$1.1 times the FWHM 
of the stellar cross-correlation function (CCF).
In the CCFs obtained from the sky-illuminated fiber B the Moon peak is visible and the continuum level
is about 2\% of the stellar CCF continuum. We corrected for the Moon light contamination by subtracting the fiber B CCF from the
fiber A CCF, and then measuring the stellar RV by means of a gaussian fit to the CCF difference.

Nineteen additional spectra were taken ($T_{exp}$=15-20 minutes, S/N$\sim$30) in previous
and following nights, spanning a time interval of about 3 years.
We estimate that the level of flux contamination in the object fiber from the fainter stellar companion is always well below $10^{-4}$.
A log of the transit observations is reported in Table\,\ref{t:logspec} and all the RVs are provided in  Table\,\ref{t:HATP20-RV-data}.

\paragraph{\bf{Qatar-2.}}
A time series of 17 spectra was obtained on 2014 April 27-28 covering a full
transit of Qatar-2\,b. The exposure time was of 15 minutes, 
resulting in spectra with S/N$\sim$6 (per pixel in 1-D spectra at 5500 \AA)
and an average RV error of $\sim$30 \ms. 
A log of the transit observations is reported in Table\,\ref{t:logspec} and the RVs are shown in Table\,\ref{t:Qatar2-RV-data}.
We note that the in-transit measurements were taken at lower and nearly constant airmass, while the initial and final 
out-of-transit data points present a relatively wide air-mass excursion.  


\begin{table*}
\caption{Log of HARPS-N observations of the planetary transits.}             
\label{t:logspec}      
\centering          
\begin{tabular}{l l c c r r c c c}     
\hline\hline       
Object  & Date\tablefootmark{(a)} & UT Start & UT End   &   $N_{\rm obs}$  & $T_{\rm exp}$[s] &   Airmass\tablefootmark{(c)}  &  Moon\tablefootmark{(b)}            & 2$^{\circ}$ fiber \\
\hline 
WASP-43  & 2013-03-11   &   21:46    &  02:01   &   32  &   450  &  1.59$\rightarrow$1.28$\rightarrow$1.46  &  NO             &    Sky            \\
HAT-P-20 & 2014-03-11   &   21:41    &  01:41   &   23  &   600  &  1.01$\rightarrow$1.85                   &  81\%/11$^{\circ}$   &    Sky            \\
Qatar-2  & 2014-04-27   &   22:08    &  02:30   &   17  &   900  &  1.51$\rightarrow$1.23$\rightarrow$1.41  &  NO             &    Sky            \\
\hline 
\end{tabular}
\tablefoot{ 
\tablefoottext{a}{Dates refer to the beginning of the night.}
\tablefoottext{b}{Fraction of illumination and angular distance from the target.}
\tablefoottext{c}{Values at first$\rightarrow$last exposure, or first$\rightarrow$meridian$\rightarrow$last exposure.}
}
\end{table*}

\subsection{Photometric data}\label{SubSec:PhRed}

A total of seven new transit light curves are presented in this study,
which were acquired with 5 different instruments.
A log of the photometric observations is reported in Table\,\ref{t:logphot}.
Following is a description, case-by-case, of the data acquisition processes and the data reduction techniques.

\paragraph{\bf{WASP-43.}}

A complete transit of WASP-43\,b was observed on November 25-26 2011 with the Copernico 1.82m telescope, 
at the Asiago Astrophysical Observatory in northern Italy. The weather conditions were perfect. 
The 886-frame photometric series, having a constant exposure time of 8~s and a net sampling cadence of about 10~s, 
was acquired with the AFOSC instrument (Asiago Faint Object Spectrograph and Camera) through a Cousins $R$ filter. 
The PSF was intentionally defocused to about 8 arcsec FWHM. The images were bias/dark subtracted and flat-field corrected 
using standard techniques. The transit light curve of WASP-43\,b was extracted by STARSKY, an independent, 
customized software pipeline to perform differential aperture photometry over defocused images 
\citep{2011A&A...527A..85N,2013A&A...549A..30N}.
The output light curves from STARSKY are automatically 
normalized by fitting a linear function to the off-transit continuum.

A second complete transit of WASP-43\,b was observed on April 15-16 2013 with the CAMELOT 
imaging camera ($R$ filter) mounted on the IAC80 telescope, at the Teide Astronomical Observatory on the Tenerife island (Spain). 
The exposure time was set to a constant 30~s, resulting in 198 full-frame images and 56~s of net cadence. 
The PSF was defocused to about 10 pixel FWHM. The images were corrected for bias and flat-field using 
standard techniques and then processed by STARSKY to get the final, differential light curve.

Another complete transit of WASP-43\,b was observed on April 19-20 2013, using the Danish 1.54m
telescope at ESO La Silla, Chile, the DFOSC imager, and a Cousins $R$
filter. The telescope was operated out of focus (see \cite{2009MNRAS.396.1023S}
 for details of the strategy and its application to
this telescope and instrument), and the observations were curtailed
immediately after egress in order to capture a time-critical event on
another target. The data were reduced using the {\sc defot} pipeline
(\cite{2014MNRAS.444..776S} and references therein), including calibration
through master bias and flat-field frames and instrumental flux
measurements by aperture photometry.
An ensemble comparison star was created from the four good comparison
stars in the images, and a linear function of time was applied to
rectify the light curve to unit flux outside transit. The weights of
the comparison stars, and the coefficients of the linear function,
were simultaneously fitted to minimise the scatter in the data outside
transit. The resulting light curve has a very low scatter of 0.57 mmag
and shows a possible starspot crossing around orbital phase
0.008.

A further complete transit of WASP-43\,b was observed with an R filter on November 11-12, 2015 within the 
EXORAP\footnote{EXOplanetary systems Robotic APT2 Photometry}  program carried out 
at the M. G. Fracastoro Station of the INAF-Catania Astrophysical Observatory with a 
80cm f/8 Ritchey-Chretien robotic telescope (APT2),  located at Serra la Nave 
(+14.973$^{\circ}$E, +37.692$^{\circ}$N, 1725 m a.s.l.) on Mt. Etna, Italy.  
The telescope is equipped with a set of standard Johnson-Cousins $UBVRI$ filters, and
an ASPEN camera with a 2k$\times$2k e2v CCD 230-42 detector that we operated with a 
binning factor of 2 (pixel scale 0.94$"$). 
Data reduction considered overscan, bias, dark subtraction and flat fielding  with the IRAF procedures 
by using the reduction pipeline specifically developed for the APT2. The night was not 
photometric and several frames were removed due to clouds after visual inspection. 
Fluxes were  extracted  by aperture photometry as implemented 
in the IDL routine {\em aper.pro}. We chose an ensemble of the three least variable
stars close to WASP-43  to get its differential photometry.
The light curve was normalized to unit flux 
dividing it by a linear best-fit function of the data outside transit.

\paragraph{\bf{HAT-P-20.}}

A nearly complete transit of HAT-P-20\,b was observed on January 16-17, 2012 with the CAMELOT camera at the IAC80 telescope through an R filter. 
The sky was perfectly clear, but a software problem forced the observer to stop the photometric series just 10~min before 
the last contact of the transit. The camera was set to read-out only one third of the available frame, to minimize the dead time between exposures. 
The exposure time was set to a constant 15~s, resulting in 483 images and 21~s of net cadence. 
The PSF was defocused to about 5 pixel FWHM. The images were corrected for bias and flat-field using 
standard techniques and then processed by STARSKY to get the final, differential light curve.

A complete transit of HAT-P-20\,b was observed on October 24-25, 2014
using the Zeiss 1.23\,m telescope at  Observatory of Calar Alto,
Spain, through a Cousins-$I$
filter. The telescope was operated out of focus, and the data were
reduced using the {\sc defot} pipeline. Bias and flat-field
calibrations were considered but not used as they had a negligible
effect on the results except for a slight increase in shot noise. An
ensemble comparison star was made from the four good comparison stars
in the images, and a quadratic polynomial versus time was applied to
rectify the light curve to unit flux outside transit. The weights of
the comparison stars, and the coefficients of the polynomial, were
simultaneously fitted to minimise the scatter in the data outside
transit.

HAT-P-20 presents a complication because of a nearby star,
originally noticed in \citet{2011ApJ...742..116B}.  \citet{2015A&A...579A.129W} found
it to lie at a separation of $6.925 \pm 0.012$ arcsec and be fainter
than HAT-P-20 in the Gunn $i$ and $z$ bands by $\Delta i = 2.01 \pm
0.08$ and $\Delta z = 1.67 \pm 0.08$ mag. The point spread function (PSF) of
this star partially overlaps that of HAT-P-20. By measuring the PSFs 
of the two stars, we found that the part of the PSF
of the fainter companion within the software aperture
for HAT-P-20 produces $18 \pm 6$\% of the flux of HAT-P-20 in the
passband we used for the observations. The  light curve was renormalized
to correct for the contaminating light from the companion.

\paragraph{\bf{Qatar-2.}}

A transit of Qatar-2\,b was observed with the IAC80 telescope at the Teide
Observatory on April 27-28, 2014,  simultaneously to the spectroscopic observations
at the TNG. We acquired a series of 133 slightly defocussed frames with the CAMELOT camera in
the R-band. Observations were affected by malfunctioning of the automatic dome tracking
which caused severe vignetting of the images. We corrected the science frames for bias and flat field
and then extracted relative photometry of our target. We selected the star and sky apertures
as well as the set of comparison stars that minimized the scatter in the data outside
transit.

\vspace{0.1cm}

For all the light curves, the timestamps were converted into the BJD(TDB) timescale using
subroutines provided by \citet{2010PASP..122..935E}.


\begin{table*}
\caption{Log of photometric observations}             
\label{t:logphot}      
\centering    
\resizebox{\textwidth}{!} {
\begin{tabular}{c c c c c c r r r c c}     
\hline\hline       
Object   &  Instr./Tel.   & Filter & Date\tablefootmark{(a)} & UT Start   & UT End   &   $N_{\rm obs}$  & $T_{\rm exp}$[s] & $T_{\rm cad}$[s]\tablefootmark{(d)} &   Airmass\tablefootmark{(c)}                                &  Moon\tablefootmark{(b)} \\
\hline 
WASP-43    & AFOSC@1.82\,m          & $R$    & 2011-11-25            &   02:57    &  05:27   &   886            &   8 &     10            & 2.28$\rightarrow$1.79                      &  NO   \\                   
   ''      & CAMELOT@IAC80          & $R$    & 2013-04-15            &   21:30    &  00:44   &   198            &   30             &     56           &   1.29$\rightarrow$1.26 $\rightarrow$1.84                    & 28\%/72$^{\circ}$    \\
   ''      & DFOSC@Danish 1.5\,m    & $R$    & 2013-04-19            &   00:02    &  01:29   &    49            &   100 &    108  &  1.10$\rightarrow$1.06$\rightarrow$1.07 & 67\%/26$^{\circ}$\\
   ''      & CCD Camera@APT2 0.8\,m & $R$    & 2015-11-12            &   02:09    &  04:54   &    73            &   120            &  128       &  2.92$\rightarrow$1.52                   &  NO \\
HAT-P-20   & CAMELOT@IAC80          & $R$    & 2012-01-16            &   22:29    &  01:19   &   483            &   15               &   21       &    1.19$\rightarrow$1.00$\rightarrow$1.01                                      &  NO   \\
   ''      &  CCD Camera@CA 1.23\,m & $I$    & 2014-10-24            &   02:02    &  05:33   &   109            &   100 &  111          &   1.54$\rightarrow$1.02                                       & 2\%/116$^{\circ}$    \\  
Qatar-2    & CAMELOT@IAC80          & $R$    & 2014-04-27            &   22:33    &  02:51   &   133            &   90             &    116            &  1.43$\rightarrow$1.22$\rightarrow$1.49  &  NO   \\

\hline 
\end{tabular}
}
\tablefoot{ 
\tablefoottext{a}{Dates refer to the beginning of the night.}
\tablefoottext{b}{Fraction of illumination and angular distance from the target.}
\tablefoottext{c}{Values at first$\rightarrow$last exposure, or first$\rightarrow$meridian$\rightarrow$last exposure.}
\tablefoottext{d}{Frame acquisition cadence.}
}
\end{table*}

%

\section{Stellar parameters} \label{Sec:Starpar}

The weighted means of all HARPS-N spectra available for the three targets were used to derive their stellar parameters. 
In particular, the equivalent widths (EWs) of iron lines taken from the list by \citet{2012MNRAS.427.2905B}, together with the {\it abfind} 
driver of the MOOG code (\cite{1973ApJ...184..839S}, version 2013), were used to obtain effective temperature ($T_{\rm eff}$), 
surface gravity ($\log g$), microturbulence velocity ($\xi_{\rm mic}$), and iron abundance ([Fe/H]). This was done by imposing the 
independence of the iron abundance on the line excitation potentials (for $T_{\rm eff}$) and EWs (for $\xi_{\rm mic}$), and the ionization 
equilibrium between \ion{Fe}{i} and \ion{Fe}{ii} (for $\log g$). 
The macroturbulence velocity ($\xi_{\rm mac}$) was fixed to the value obtained using the \citet{2005ApJS..159..141V} relationship depending on $T_{\rm eff}$ and $\log g$.
After fixing the stellar parameters at the values derived through 
the EWs, a spectral synthesis was performed using the {\it synth} driver of the same code to measure the projected rotational 
velocity ($V\sin{I_\star}$) and following the prescriptions given by \citet{2011A&A...526A.103D}. All the analysis was performed differentially 
with respect to the Sun, thanks to a mean Vesta spectrum acquired with HARPS-N. 

For further details on the procedures based on EWs and spectral synthesis, we refer to the aforementioned papers, together with 
other works within the GAPS project (see, e.g., \cite{2013A&A...554A..28C, 2015A&A...575A.111D, 2016A&A...588A.118M}). Results of the 
spectroscopic analysis here applied to determine the stellar parameters are listed in Table~\ref{t:bestfitpar}.

We used the average spectra also to analyse the CaII H\&K lines and measure the chromospheric Mount Wilson S-index for WASP-43 and HAT-P-20,
while for Qatar-2 the S/N was too low to derive a reliable measurement.
We then calculated the log($R'_{\rm HK}$) index, following the prescriptions in  \citet{2015MNRAS.452.2745S}
\footnote{We did not use the HARPS-N DRS for the measurement of the S-index because  its application to average spectra is not straightforward. Also,
since the B$-$V colors of our targets are close to 1.2, for the calculation of the log($R'_{\rm HK}$) index, we preferred to follow \citet{2015MNRAS.452.2745S} rather than 
\citet{2011arXiv1107.5325L}}.
We used our determinations of the $T_{\rm eff}$ and the  empirical calibrations reported in \citet{1996ApJ...469..355F}
to derive the needed intrinsic B$-$V colour index. We obtain  B$-$V = 1.19$\pm$0.06 and 1.135$\pm$0.025
for WASP-43 and HAT-P-20, respectively. We notice that the observed B$-$V colors (see Table~\ref{t:bestfitpar}) are slightly larger
than the intrinsic ones, possibly due to interstellar reddening.
The values of the S and  log($R'_{\rm HK}$) indices are reported in Table~\ref{t:bestfitpar}.

%

\section{Light curves and radial velocities analysis} \label{Sec:LCRV}

%
We have developed a code, within the MATLAB software ambient 
\footnote{MATLAB R2015b, Optimization Toolbox 7.3 and Curve Fitting Toolbox 3.5.2, The MathWorks, Inc., Natick, Massachusetts, United States.}, 
for modelling and fitting planet transits observations. The code can simultaneously fit any number of (in- and out-transit) RV data sets,
as well as transit light curves in different filters.

The model considers the parameters necessary
to fully describe the planet and star position and velocity vectors
at any given time, i.e: the masses of the star $M_\star$ and of the planet $M_{\rm p}$,
the orbital period $P$ and eccentricity $e$, the epoch $\tau$ and argument $\omega$ of
periastron, the systemic RV $\gamma$; the orbital space orientation is
described by the inclination angle $i_p$ and the misalignment angle $\lambda$.
The third angle, the longitude of the ascending node, is not considered as it does
not affect the RV and photometric measurements.
Other parameters necessary to model the RM effect and the light curves are the radii of the star $R_\star$
and of the planet $R_{\rm p}$, the stellar projected rotational velocity $V\sin{I_\star}$,
and the limb-darkening coefficients. Our model implements each of the 5 equations
(linear, quadratic, root-square, logarithmic, and a 4-coefficient law) proposed
by \citet{2011A&A...529A..75C} to describe the limb-darkening law.
Other effects that can affect the measurements, like 
stellar surface inhomogeneities (spots, faculae, etc.), stellar differential
rotation and convective blue-shift, are not included in the model.

For the analysis of the RM effect we implemented a numerical model based on the following assumptions.
We tried to reproduce the observed CCF by modelling an average photospheric line profile.
The stellar disc is sampled by a matrix of 1000$\times$1000 elements, each element being represented by 
a Gaussian line profile with a given width $\sigma_{el}$, Doppler-shifted according to the stellar rotation, 
and weighted by appropriate limb-darkening coefficients.
$\sigma_{el}$ is also a parameter of the fit, however we always fix it, as it turned out that the model RVs are largely insensitive to its exact value. 
The value of $\sigma_{el}$ is chosen by adding in quadrature the values of $\xi_{\rm mic}$ and $\xi_{\rm mac}$.
The resulting line profile is then convolved by the instrumental profile of HARPS-N, 
assumed Gaussian with $\sigma_{IP}=1.108$\,km\,s$^{-1}$. 
The model also takes into account the actual area of the stellar photospheric disc occulted and the smearing 
due to the planet's displacement during an exposure.  
The corresponding RV shift is then computed by a Gaussian fit of this resulting line profile, 
analogously to the HARPS-N DRS.
The same numerical approach is also used for the modelling and analysis of the light curves. Limb darkening
coefficients can be fitted independently for each light curve.

The best-fitting model to the data is found
via a sigma-weighted, robust least-squares minimization. 
The region of the
parameters space to be explored can be limited
providing upper and lower limits to the parameter values.
Most importantly, any number of linear and non-linear constraints
can be set: this allows to place limits to other parameters
(such as $K$, $T_{14}$ , $b$, etc.), even though they are not direct parameters
of the fit.
The mass of the star is preliminarily determined from 
evolutionary track models, adopting the values of the atmospheric parameters
determined previously and using the $a/R_\star$ value derived from an independent fit of the light curve.
The uncertainties on the best-fit values are obtained by means of a bootstrap algorithm.

We use our code to analyse all the RVs and light curves presented in this work.
Depending on the type and quality of our data as compared to those in literature, we decide 
whether a parameter is to be fitted or fixed to the 
value available in literature.

\subsection{WASP-43}\label{SubSec:DataAnalysis:WASP43}

To begin with, we analysed
the 4 transit light curves (LCs) taken individually.
The best-fit values of the relevant parameters are reported in Table\,\ref{t:fitLC_wasp43}.
Following \citet{2007ApJ...664.1190S}, we used the $a/R_\star$ values to derive the
stellar density, resulting in a sigma-clipped weighted average of $\rho_*$=2.47 $\pm$ 0.11 $\rho_{\odot}$.
The sigma-clipping effectively excluded the APT2 value that is off by >2$\sigma$ with respect to the other three.        
The stellar density, together with our determinations of $T_{\rm eff}$ and [Fe/H],
was used to estimate the mass of the star by comparison with 
the Yonsei-Yale evolutionary tracks \citep{2004ApJS..155..667D}.
Following \citet{2011MNRAS.417.2166S}, we accounted for systematic uncertainties in the stellar models
by adding an extra 5\% to the formal errors, and obtained
$M_\star=0.688 \pm 0.037$ M$_\odot$.

Our determinations of the 4 epochs of mid-transit are compatible with 
the most recently published ephemerides  
\citep{2016AJ....151..137H,2016AJ....151...17J,2015PASP..127..143R,2014A&A...563A..40C}.
In the following analyses, we fixed the orbital period to the value reported in 
\citet{2016AJ....151..137H}, where all the LCs available in literature were analysed homogeneously: P = 0.813473978 $\pm$ 3.5$\times$10$^{-8}$ days.
%

\begin{table*}
\caption{
Results of the individual analyses of the four WASP-43\,b and two HAT-P-20\,b transit light curves. The columns report the following parameters: $a/R_\star$ is
the ratio between the semi-major orbital axis and the stellar radius; $R_{\rm p}/R_\star$ is the planet to star radii ratio; $i\, [^{\circ}]$ is the orbital inclination angle;
$b$ is the transit impact parameter; $T_{14}$ is the transit duration;  $T_{\rm C}$ is the epoch of mid-transit; $u$ is the coefficient of the linear limb-darkening law according to
Equation (1) in \citet{2011A&A...529A..75C}. 
}
\label{t:fitLC_wasp43}
\centering
\resizebox{\textwidth}{!} {
\begin{tabular}{lcccccc}
\hline
\hline

Instr./Tel.                     & $a/R_\star$        & $R_{\rm p}/R_\star$  & $i_{\rm p}\, [^{\circ}]$ &  $T_{14} $ [days]      & $T_{\rm C} $ [BJD$-$2400000]          & $u$ \\

\hline
\multicolumn{7}{c}{ {\bf WASP-43} } \\
\hline
AFOSC@1.82m                     & 4.896 $\pm$ 0.084  & 0.1590  $\pm$ 0.0017  & 82.27 $\pm$ 0.29 & 0.0512  $\pm$ 0.0010   & 55891.67847  $\pm$ 0.00018   & 0.66 $\pm$ 0.13   \\ 
CAMELOT@IAC80                   & 4.989 $\pm$ 0.070  & 0.1594  $\pm$ 0.0016  & 82.12 $\pm$ 0.21 & 0.04943 $\pm$ 0.00073  & 56398.47228  $\pm$ 0.00016   & 0.511 $\pm$ 0.075 \\
DFOSC@Danish 1.5m               & 5.039 $\pm$ 0.052  & 0.15538 $\pm$ 0.00086 & 82.03 $\pm$ 0.19 & 0.04792 $\pm$ 0.00036  & 56402.539119 $\pm$ 0.000088  & 0.484 $\pm$ 0.057 \\
CCD Camera@APT2 0.8\,m          & 4.737 $\pm$ 0.090  & 0.1702  $\pm$ 0.0021  & 81.36 $\pm$ 0.48 & 0.05156 $\pm$ 0.00030  & 57339.66239  $\pm$ 0.00011   & 0.54  $\pm$ 0.11  \\
\hline
\hline 
\multicolumn{7}{c}{ {\bf HAT-P-20} } \\
\hline
CAMELOT@IAC80                   & 10.74 $\pm$ 0.44   & 0.1534 $\pm$  0.0034 & 86.32 $\pm$  0.40 & 0.0796 $\pm$  0.0016   & 55943.52369 $\pm$   0.00025      &  0.51 $\pm$ 0.12\\
CCD Camera@CA 1.23\,m           & 11.48 $\pm$ 0.33   & 0.1542 $\pm$  0.0027 & 87.00 $\pm$  0.26 & 0.0791 $\pm$  0.00048  & 56955.634275 $\pm$  0.000055     & 0.675 $\pm$  0.068 \\
\hline
\hline 
\end{tabular}
}
\end{table*}


We then analysed the out-of-transit RVs.
In addition to our 22 measurements, we also considered the 23 CORALIE RV values
reported in \citet{2011A&A...535L...7H} and \citet{2012A&A...542A...4G}, allowing for a 
constant RV offset between the two data sets. We found that the best-fit circular and eccentric
orbits were virtually indistinguishable from each other, therefore we opted for fixing $e$=0.
 Strong constraints on the eccentricity  were already put by \citet{2012A&A...542A...4G} (0.0035$^{+0.0060}_{-0.0025}$)
and \citet{2014ApJ...781..116B} (0.010$^{+0.010}_{-0.007}$),
based also on the timing of secondary eclipse. 
From the best-fit value of the RV curve semi-amplitude $K$ = 551 $\pm$ 8 \ms,
and the orbital inclination angle that we derived from the LCs analysis (see Table\,\ref{t:fitLC_wasp43}), 
we calculated the planetary mass.

The next step was to analyse the RV time-series covering the planetary transit and to measure
the RM effect. By fixing the other relevant parameters, as obtained by the LCs and out-of-transit RVs fits,
we derived a best-fit value for the sky-projected spin-orbit angle $\lambda$ and 
stellar rotational velocity $V\sin{I_\star}$.

\begin{figure}
  \resizebox{\hsize}{!}{\includegraphics{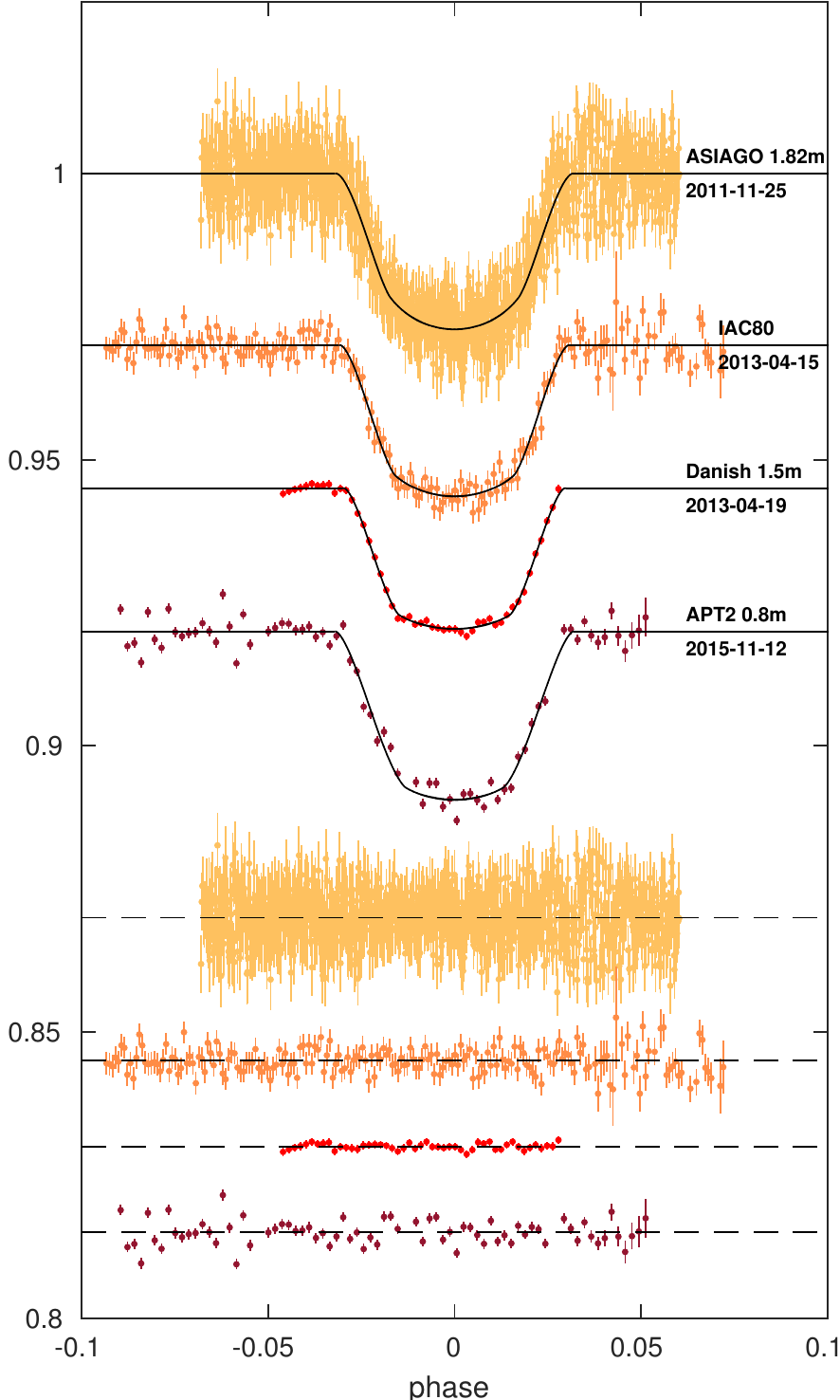}}
  \caption{WASP-43 normalized light curves with best-fit models superimposed. For clarity a vertical offset was applied to the different 
  LCs. The lower part of the diagram shows the best-fit residuals. Data are phase-folded according to the orbital period reported in 
  Table \ref{t:bestfitpar} and the epochs of mid-transit reported in Table \ref{t:fitLC_wasp43} . For each LC we report
  the telescope and the date of observation. In all cases a Johnson R filter was used. 
  }
  \label{Fig:WASP43_LC}
\end{figure}

\begin{figure}
  \resizebox{\hsize}{!}{\includegraphics{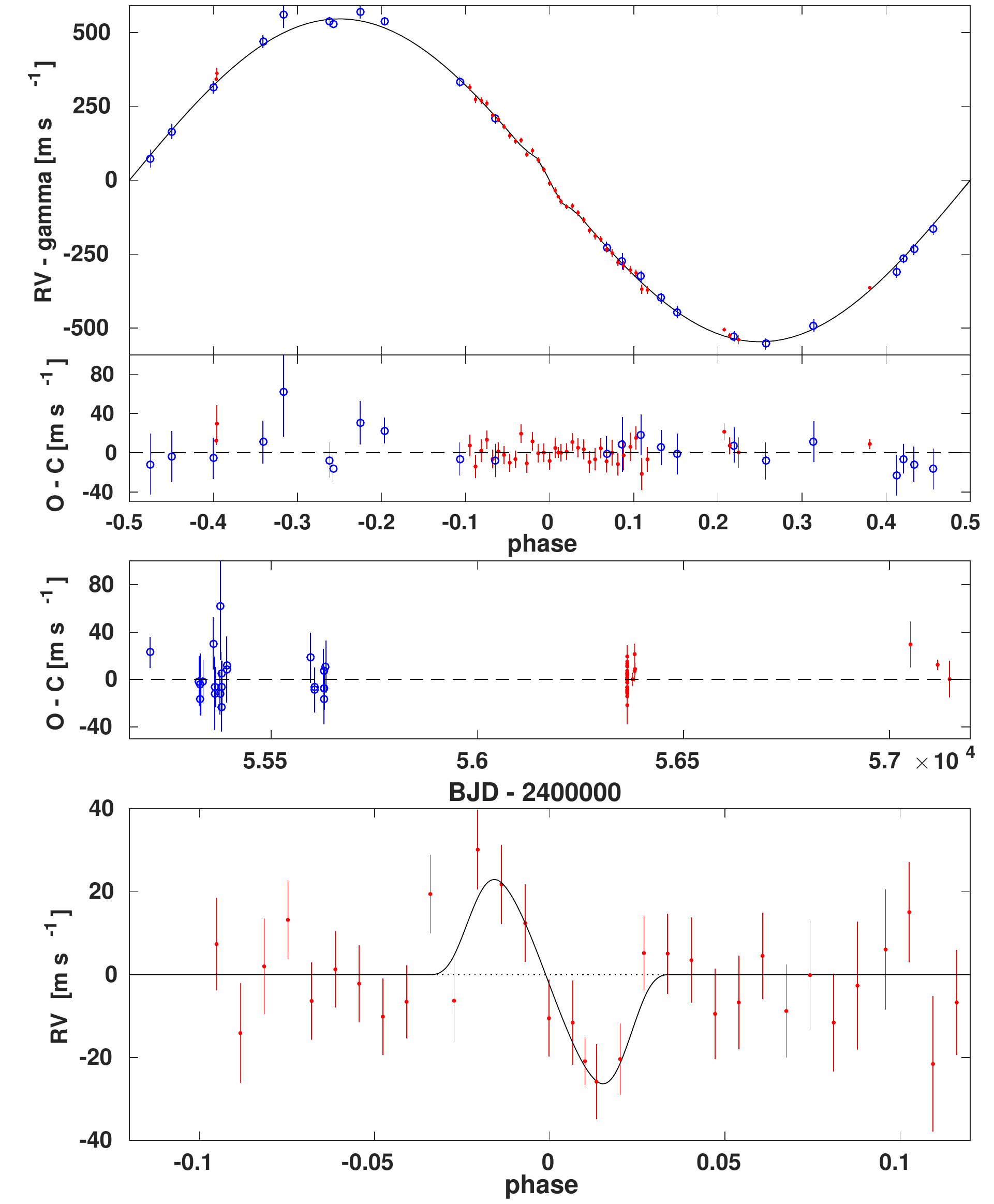}}
  \caption{WASP-43 radial velocities. Red dots represent HARPS-N data while blue open circles are CORALIE data.
            \textit{Top panels:} phase-folded RVs with the best-fit RV curve superimposed, and below the corresponding residuals. 
           \textit{Middle panel:} RV residuals plotted as a function of the Barycentric Julian Date (BJD) show
                                  no evidence of a long term RV trend.
           \textit{Bottom panel:} Zoom in the RV time-series covering the transit. In order to highlight the RM 
                                  effect, the orbital RV trend was subtracted. 
  }
  \label{Fig:WASP43_RV}
\end{figure}

Finally, using as a first guess the values obtained from the previous analyses,  
we made a global joint fit of all the RVs and LCs in order to derive a fully consistent set of best-fit
orbital and physical parameters. The final results are reported in Table\,\ref{t:bestfitpar}.
Aside from our new determination of $\lambda$ = 3.5$\pm$6.8 deg, the values of the stellar and planet
parameters are consistent with previous measurements \citep{2011A&A...535L...7H,2012A&A...542A...4G,2014A&A...563A..40C}.
The four phase-folded LCs and their best-fit models are displayed in Fig. \ref{Fig:WASP43_LC}.
The RV measurements with the orbital and RM best-fit curves are shown in Fig. \ref{Fig:WASP43_RV}.

By combining our measurements of $R_\star$ and V$\sin{I_\star}$ with the stellar rotational
period $P_{\rm rot}$ = 15.6$\pm$0.4 days reported in \citet{2011A&A...535L...7H},
we can estimate (see formula (8) in \cite{2007AJ....133.1828W}) the value of $\sin{I_\star}$ to be 1.08$\pm$0.25. Being  $\sin{I_\star}$>1
physically impossible, we deduce that the true value of $\sin{I_\star}$ must be very close
to 1. Therefore it is $I_\star$$\simeq$90 deg, in fact we derive that $I_\star$>72 deg with a 
68\% level of confidence.
Similarly (see formula (7) in \cite{2007AJ....133.1828W}), we derive that the  
true orbital misalignment angle $\varPsi$ 
is <20 deg with a 68\% level of confidence.


\begin{table*}
\caption{
Planetary and stellar parameters for the three systems here studied.
}
\label{t:bestfitpar}
\centering
\resizebox{\textwidth}{!} {
\begin{tabular}{lccc}
\hline
\hline
Parameter [Unit]                       &  WASP-43                         &  HAT-P-20                    & Qatar-2 \\
\hline
B [mag]\tablefootmark{(j)}                                &  13.796 $\pm$ 0.022              &  12.539 $\pm$ 0.075          & 14.582 $\pm$ 0.022        \\
V [mag]\tablefootmark{(j)}                                &  12.464 $\pm$ 0.028              &  11.339 $\pm$ 0.031          & 13.417 $\pm$ 0.023        \\
J [mag]\tablefootmark{(k)}                                &   9.995 $\pm$ 0.024              &   9.276 $\pm$ 0.022          & 11.350 $\pm$ 0.026        \\
H [mag]\tablefootmark{(k)}                                &   9.397 $\pm$ 0.025              &   8.743 $\pm$ 0.021          & 10.794 $\pm$ 0.022        \\
K [mag]\tablefootmark{(k)}                                &   9.267 $\pm$ 0.026              &   8.601 $\pm$ 0.019          & 10.619 $\pm$ 0.021        \\
Space velocity ($U$,$V$,$W$) [\kms]                       &  (-2.7,\,-10.8,\,-20.4)          &  (19.1,\,-27.7,\,-21.4)      &  ---                      \\
\multicolumn{4}{c}{ {\it Stellar spectra characterization} } \\
Effective temperature, $T_{\rm eff}$ [K]   &  4500$\pm$100                &  4595$\pm$45            & 4640$\pm$65           \\
Surface gravity, $\log g_\star$ [\cmss]    &  4.50$\pm$0.20               &  4.52$\pm$0.09          & 4.51$\pm$0.12         \\
Iron abundance, [Fe/H]                     &  $-$0.01$\pm$0.15            &  0.22$\pm$0.09          & 0.13$\pm$0.10         \\
Microturbulence vel. , $\xi_{\rm mic}$ [\kms]        &  1.00$\pm$0.15               &  0.74$\pm$0.27          & 0.90$\pm$0.35         \\
Macroturbulence vel. , $\xi_{\rm mac}$ [\kms] \tablefootmark{(l)}         &  2.03               &  2.17          &   2.24                                \\

Proj. rot. vel., $V\sin{I_\star}$ [\kms]   &  2.6$\pm$0.5                 &  2.0$\pm$0.5            & 2.0$\pm$1.0           \\
<MW S-index>                               &  1.647$\pm$0.059             &  1.20$\pm$0.13          &  ---                  \\
<log($R'_{\rm HK}$)>                       &  -4.35$\pm$0.10              &  $-$4.40$\pm$0.06       &  ---                  \\
\hline
\multicolumn{4}{c}{ {\it RV and photometric data fit} } \\
Star mass, $M_\star$ [M$_\odot$]           &  0.688 $\pm$ 0.037            &  0.742 $\pm$ 0.042          & 0.798 $\pm$ 0.040         \\ 
Planet mass, $M_{\rm p}$ [M$_{\rm J}$]     &  1.998 $\pm$ 0.079            &  7.22$\pm$ 0.36            & 2.616 $\pm$ 0.071         \\    
Star radius, $R_\star$ [R$_\odot$]         &  0.6506 $\pm$  0.0054         &  0.6796 $\pm$  0.0054       & 0.793 $\pm$ 0.024         \\ 
Planet radius, $R_{\rm p}$ [R$_{\rm J}$]   &  1.006 $\pm$ 0.017            &  1.025 $\pm$ 0.053          & 1.281 $\pm$ 0.039         \\ 
Orbital period, $P$ [days]                 &  0.813473978$\pm$0.000000035\tablefootmark{(a)}  &  2.875316938$\pm$0.00000019      & 1.33711647 $\pm$ 0.00000026 \tablefootmark{(f)}   \\
Eccentricity, $e$                          &  0 (fixed)                    &  0.0172  $\pm$ 0.0016       &  0 (fixed)       \\
Longitude of periastron, $\omega$ [deg]    &  90 (fixed)                   &  342.7  $\pm$ 7.3           &  90 (fixed)      \\
Orbital inclination, $i_{\rm p}$ [deg]     &  82.109  $\pm$  0.088         &  86.88  $\pm$  0.31         &  86.12 $\pm$ 0.08  \tablefootmark{(f)}     \\ 
Epoch of periastron, $\tau$ [BJD]          &  ---                          &  2455942.681 $\pm$ 0.016    & ---         \\
Barycentric RV, $\gamma$ [\ms]             &  $-$3595.5 $\pm$ 4.3 \tablefootmark{(b)} & $-$18087.44 $\pm$ 0.7 \tablefootmark{(g)}             &  -23977.5 $\pm$ 7.1       \\
Barycentric RV, $\gamma_2$ [\ms]           &  $-$3588.1 $\pm$ 2.7 \tablefootmark{(c)} &  $-$18093.36 $\pm$ 0.8\tablefootmark{(g)}       &   ---     \\
Proj. spin-orbit angle, $\lambda$ [deg]    &  3.5 $\pm$ 6.8                & -8.0 $\pm$ 6.9             & 15 $\pm$ 20        \\
Proj. rot. vel., $V\sin{I_\star}$ [\kms]   &  2.26 $\pm$ 0.54              & 1.85 $\pm$ 0.27             & 2.09 $\pm$ 0.58       \\ 
Stellar rotational period, $P_{\rm rot}$ [days] &  15.6$\pm$0.4\tablefootmark{(d)} & 14.48$\pm$0.02\tablefootmark{(e)}                    & 18.77 $\pm$ 0.29        \\
\hline
\multicolumn{4}{c}{ {\it Derived parameters} } \\
Orbital semi-major axis, $a$ [AU]                          &      0.01504 $\pm$ 0.00029      & 0.03593 $\pm$ 0.00029   &  0.02205 $\pm$ 0.00037   \\  
Transit duration, $T_{14}$ [days]                          &      0.0485   $\pm$ 0.010       & 0.07900   $\pm$ 0.00052 &  ---     \\   
Impact parameter, $b$                                      &      0.689    $\pm$ 0.013       & 0.622    $\pm$ 0.059    &  0.405 $\pm$ 0.028     \\
$a/R_\star$                                                &    4.97 $\pm$  0.14   & 11.36  $\pm$ 0.25    &  5.98 $\pm$  0.28                         \\
$R_{\rm p}/R_\star$                                        &   0.1588 $\pm$ 0.0040 & 0.155$\pm$ 0.010       & 0.166 $\pm$ 0.010      \\
Spin-orbit angle $\varPsi$ [deg]                           &      <20\tablefootmark{(i)}     & 36 $_{-12}^{+10}$        &  <43\tablefootmark{(i)}                  \\
Star incl. angle $I_\star$ [deg]                           &     >72\tablefootmark{(i)}     & 53   $\pm$ 12     &  >58\tablefootmark{(i)}                 \\
RV-curve semi-amplitude, $K$ [\ms]                         &      551.0$\pm$3.2              & 1249.5$\pm$ 1.2         & 558.7 $\pm$ 5.9 \tablefootmark{(h)}   \\
Star density, $\rho_\star$ [\gcm]                          &      2.526  $\pm$ 0.080         & 2.36  $\pm$ 0.16        & 2.240 $\pm$ 0.023 \tablefootmark{(f)}   \\
Star surface gravity, $\log g_\star$ [\cmss]               &      4.647  $\pm$ 0.011         & 4.643$\pm$ 0.020        & 4.541$\pm$ 0.048  \\
Planet density, $\rho_{\rm p}$ [\gcm]                      &      2.43 $\pm$ 0.14            & 8.31 $\pm$ 0.38         & 1.54 $\pm$ 0.20    \\
Planet surface gravity, $\log g_{\rm p}$ [\cmss]           &      3.696  $\pm$ 0.018         & 4.231  $\pm$ 0.019      & 3.597 $\pm$ 0.038      \\
Planet equilibrium temperature, $T_{\rm p}$ [K]            &      1426.7  $\pm$  8.5         & 964 $\pm$ 10         & 1342 $\pm$ 15      \\  

\hline
\end{tabular}
}
\tablefoot{ 
\tablefoottext{a}{Adopted from \citet{2016AJ....151..137H};}
\tablefoottext{b}{CORALIE data;}
\tablefoottext{c}{HARPS-N data;}
\tablefoottext{d}{Adopted from \citet{2011A&A...535L...7H};}
\tablefoottext{e}{Value adopted from \citet{2014AN....335..797G}, error derived by us with an independent re-analysis of the data;}
\tablefoottext{f}{Adopted from \citet{2014MNRAS.443.2391M};}
\tablefoottext{g}{Off-transit ($\gamma$) and in-transit ($\gamma_2$) HARPS-N data;}
\tablefoottext{h}{Adopted from \citet{2012ApJ...750...84B};}
\tablefoottext{i}{At the 68\% level of confidence.}
\tablefoottext{j}{APASS catalogue \citep{2016yCat.2336....0H}.}
\tablefoottext{k}{2MASS catalogue \citep{2003yCat.2246....0C}.}
\tablefoottext{l}{Fixed to the values obtained following \citet{2005ApJS..159..141V}}
}
\end{table*}



\subsection{HAT-P-20}\label{SubSec:DataAnalysis:HATP20}

\begin{figure}
  \resizebox{\hsize}{!}{\includegraphics{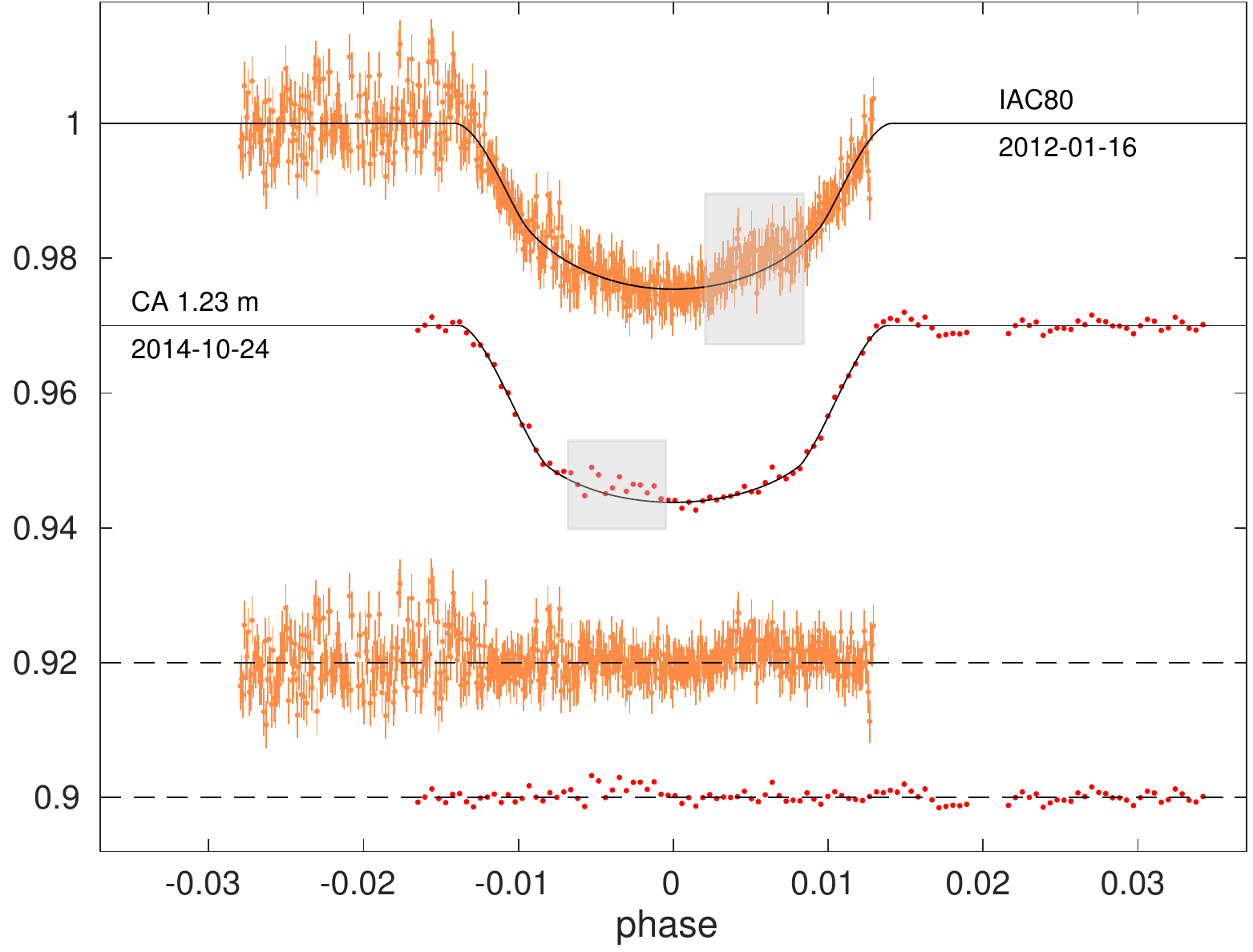}}
  \caption{HAT-P-20 light curves (top) and best-fit residuals (bottom). Data are phase-folded according to the best-fit period
  reported in Eq. (\ref{eph_HATP20}).
  For each LC the telescope and date of observation are indicated.
  For the IAC80 LC a Johnson-$R$ filter was used, for the CA LC a Cousins-$I$ filter. 
  Shadowed areas indicate phases of possible spot crossing.
  }
  \label{Fig:HATP20_LC}
\end{figure}

HAT-P-20 shows clear signatures of a high level of stellar activity.
\citet{2014AN....335..797G} have analysed the SuperWASP photometry of HAT-P-20 
to derive the rotational period and found a modulation with a peak-to-valley amplitude
as high as $\sim$0.04 mag. 
Both our two transit light curves show evidence of the planet crossing
a star spot (see Fig. \ref{Fig:HATP20_LC}). Also, the analysis of the RV measurements reveals jitter
at the level of $\sim$20 \ms. 

We first analysed the two LCs individually.
In Table\,\ref{t:fitLC_wasp43}, we report the values of the fitted transit parameters.
We find values for $R_{\rm p}/R_\star$ that are consistent with each other
but significantly larger than any of the values reported in \citet{2014AN....335..797G}.
A possible explanation for this discrepancy is that 
the two transits observed by us occurred during
a phase of higher stellar spots filling factor \citep{2012A&A...539A.140B}. 
The two light curves with the best-fit transit model superimposed are displayed in Fig. \ref{Fig:HATP20_LC}.
Thanks to the robust algorithm the best-fit is only marginally affected by the spot crossing regions.


\begin{figure}
  \resizebox{\hsize}{!}{\includegraphics{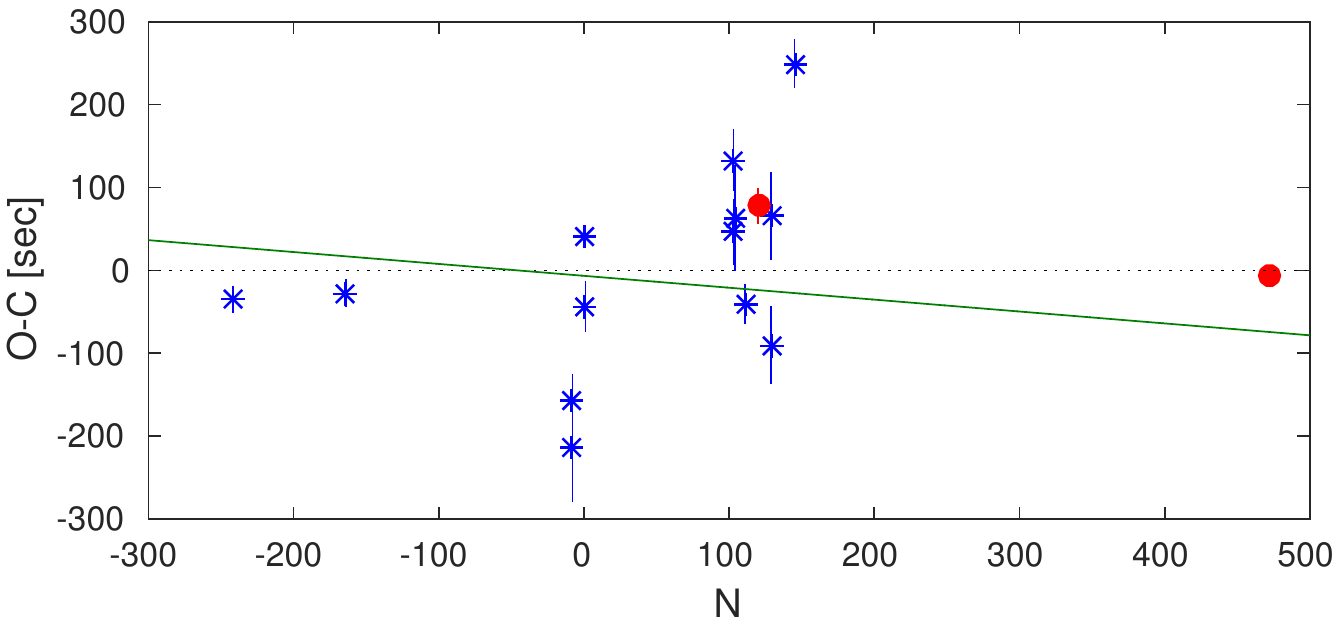}}
  \caption{O-C diagram obtained by a linear fitting (i.e constant period) of the HAT-P-20\,b mid-transit epochs.
  Blue asterisks are from \citet{2014AN....335..797G}, red dots correspond to the 
  two transit observations reported in this work.
  The green continuous line represents the ephemerides calculated in \citet{2014AN....335..797G}.  
  }
  \label{Fig:HATP20_ephemerides_fit}
\end{figure}

By adding our two determinations of the epoch of mid-transit to the
values listed in \citet{2014AN....335..797G}, we calculated new transits ephemerides.
A weighted least-square linear fit (see Fig. \ref{Fig:HATP20_ephemerides_fit}) yielded:


\begin{equation} \label{eph_HATP20}
\begin{array}{ccl}
T_0\,(\mathrm{BJD}_{\mathrm{TDB}}) & = & (2455598.484742\pm 0.000073) \,+\\ 
&& + N\,(2.87531694 \pm 0.00000019).
\end{array}
\end{equation}


 Fixing the values obtained for $T_0$ and $P$, we then performed
a global fit of the two light curves and all the HARPS-N RVs.
We treated the RV time-series on the night of the transit as an
independent data-set by adding a parameter to the fit which represents the 
barycentric radial velocity for the night of the transit.
This was necessary to account for an RV offset most likely caused
by the stellar jitters.

\begin{figure}
  \resizebox{\hsize}{!}{\includegraphics{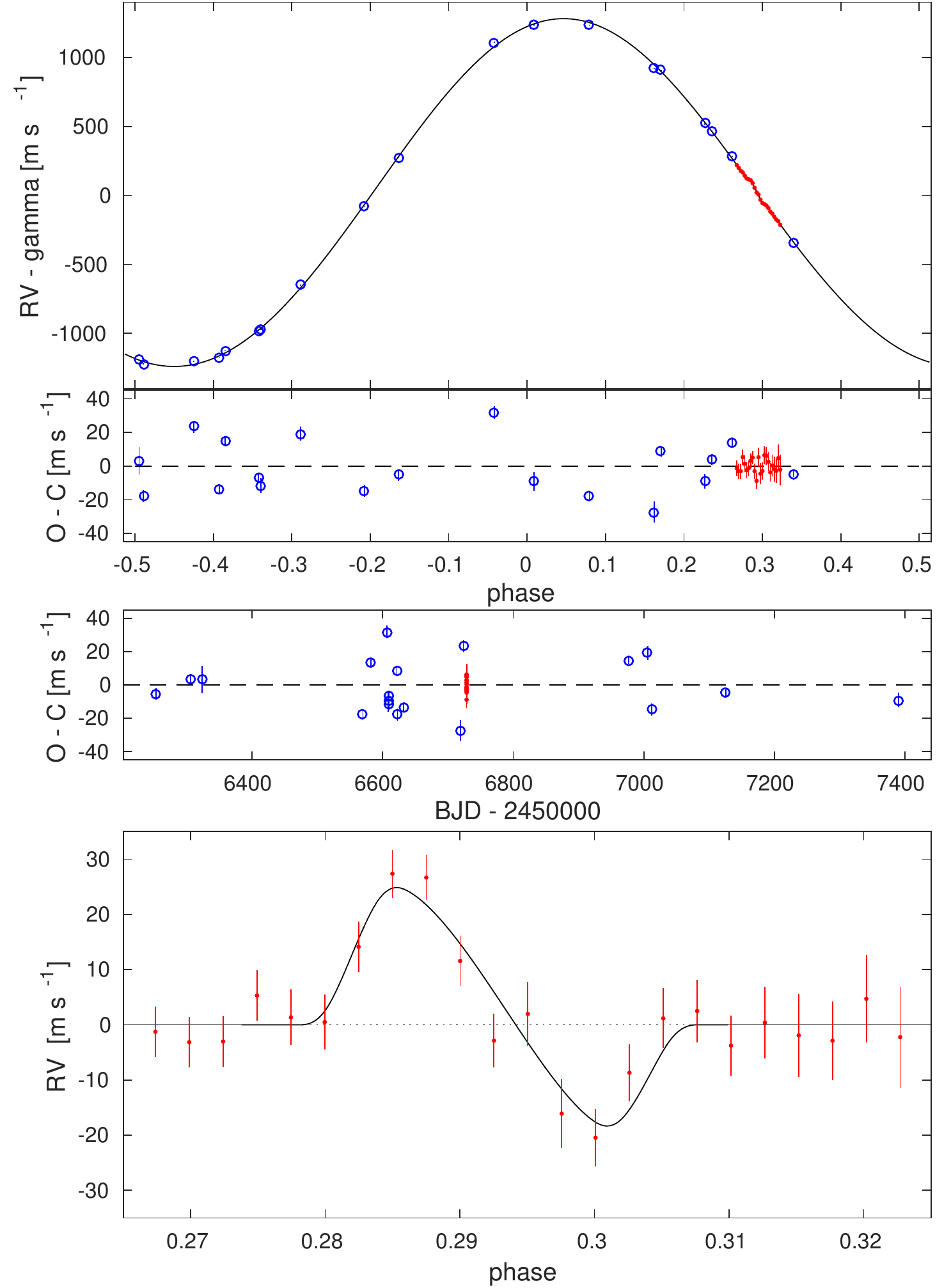}}
  \caption{HAT-P-20 radial velocities.  Data taken during the night of the transit are displayed with red dots. 
           \textit{Top panels:} phase-folded RVs with the best-fit RV curve superimposed, and below the corresponding residuals.
           \textit{Middle panel:} RV residuals plotted as a function of the Barycentric Julian Date (BJD) show
                                  no evidence of a long term RV trend.
           \textit{Bottom panel:} RV time-series covering the transit. In order to highlight the RM 
                                  effect, the orbital RV trend was subtracted.
  }
  \label{Fig:HATP20_RV}
\end{figure}

The phase-folded RVs and the best-fit model are displayed in the top panel of Fig.  \ref{Fig:HATP20_RV}.
Our data confirm that HAT-P-20\,b moves on an orbit with a small but significant
eccentricity. Our best-fit value is $e$ = 0.0172  $\pm$ 0.0016,
in agreement with previously reported values ($e$ = 0.015  $\pm$ 0.005, \cite{2011ApJ...742..116B}; 
$e$ = 0.0158$^{+0.0041}_{-0.0036}$, \cite{2014ApJ...785..126K}; $e$ = 0.0171$^{+0.0018}_{-0.0016}$, \cite{2015ApJ...805..132D}).
For the mass of the planet we find  $M_{\rm p}$ = 7.22 $\pm$ 0.36  M$_{\rm J}$, in agreement with the literature value 
($M_{\rm p}$ = 7.24 $\pm$ 0.18  M$_{\rm J}$,  \cite{2014ApJ...785..126K}).
However, given the larger planet radius ($R_{\rm p}$ = 1.025 $\pm$ 0.053 R$_{\rm J}$) that we obtain from the transit light curves analysis,
we derive a planet density significantly lower  ($\rho_{\rm p}$ = 8.31 $\pm$ 0.38 \gcm) than previously reported
(13.78 $\pm$ 1.50 \gcm, \cite{2011ApJ...742..116B}).
After subtraction of the best-fit 1-planet model, the RV residuals show a scatter (rms$\simeq$16 \ms) largely exceeding
the typical internal RV uncertainties ($\sigma\simeq$3.5 \ms) (see mid-panel in Fig. \ref{Fig:HATP20_RV}). A frequency analysis of the RV residuals does not point out any
significant periodicity, therefore we ascribe the large residuals to stellar activity induced RV jitter. However, we have checked 
that there is no evident correlation between the RV residuals and the CCF bisector span. 
In the computation of the errors on the best-fit parameters we accounted for a  RV-jitter term. 
\citet{2014ApJ...785..126K}, with 13 RV measurements spanning 1331 days, report a negative linear trend 
of $-0.0141^{+0.0073}_{-0.0078}$ m s$^{-1}$ day$^{-1}$. 
By allowing for a linear trend in our fit of HARPS-N RVs (19 out-of-transit data points spanning 1137 days), 
we obtain a positive, but not significant, trend of 7.5e-03 $\pm$ 15e-03 m s$^{-1}$ day$^{-1}$. 

The best fit model of the RM effect gives  V$\sin{I_\star}$ = 1.85 $\pm$ 0.27 \kms, slightly lower but consistent with
the spectral synthesis determination, and $\lambda$ = -8.0 $\pm$ 6.9 deg which is marginally different from zero.
Using the value of the stellar rotational
period $P_{\rm rot}$ = 14.48$\pm$0.02 days reported in \citet{2014AN....335..797G},
we can estimate the stellar spin-axis inclination to be  $I_\star$=53$\pm$12 deg, and 
the true planet orbital misalignment angle  $\varPsi$=36$_{-12}^{+10}$ deg, 
meaning that HAT-P-20\,b lies on a significantly inclined orbit.
As a note of caution on this result, we stress that, if the star 
experiences a significant differential rotation, we are underestimating 
 V$\sin{I_\star}$,
while higher values of V$\sin{I_\star}$  translate in
$I_\star$ closer to 90 deg and $\varPsi$ closer to 0 deg.

\subsection{Qatar-2}\label{SubSec:DataAnalysis:Qatar2}

\begin{figure}
  \resizebox{\hsize}{!}{\includegraphics{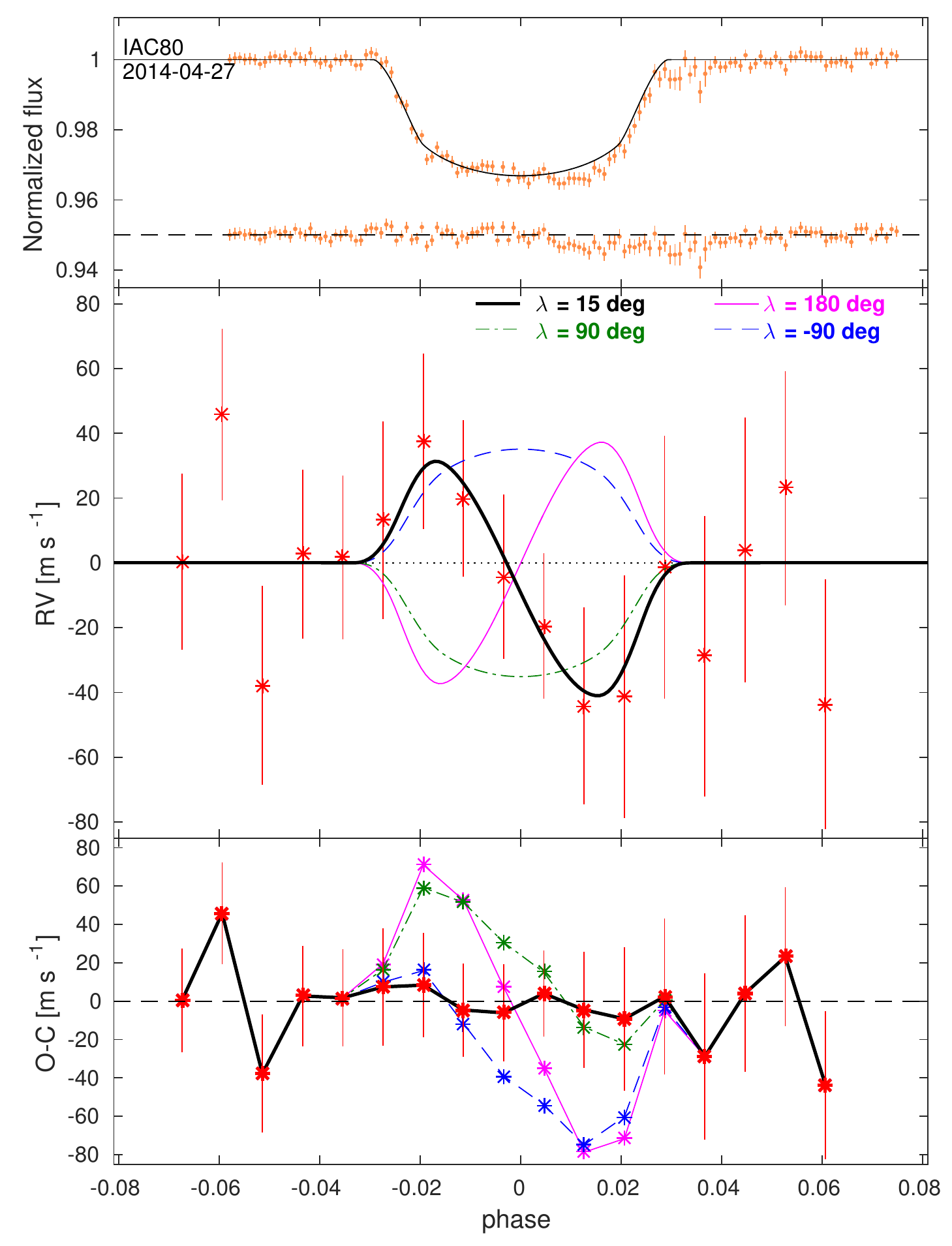}}
  \caption{Simultaneous photometric and RV monitoring of the transit of Qatar-2\,b.
  Top panel: IAC80 photometric data. The black line is the model light curve, but it 
  is not a fit to the data (see text in Sec. \ref{SubSec:DataAnalysis:Qatar2} for details).
 Middle panel: RV time-series covering the transit (red asterisks).
  In order to highlight the RM effect, the orbital RV trend was subtracted.
  Together with our best-fit model (black thick line), we show also models with 
  $V\sin{I_\star}$ = 2.0 \kms\, and $\lambda$=-90, 90, 180 deg (dashed blue, dot-dashed green, magenta lines, respectively).
  Bottom panel: RV residuals for the best-fit model of the RM effect and for the other three models shown in the middle panel. 
  }
  \label{Fig:fig_QATAR2_LC_RM_02}
\end{figure}

The Qatar-2\,b transit photometry, which we acquired simultaneously to the HARPS-N
RV monitoring, resulted to be affected by a trend
that starts shortly after the mid-transit phase and extends well beyond the end of the transit
(see the top panel in Fig. \ref{Fig:fig_QATAR2_LC_RM_02}).
We traced back the origin of this trend to image vignetting (see Section \ref{SubSec:PhRed}),
but we could not reliably model and correct for it.
Therefore, although we show the light curve, it was not actually used in our data fit.
Instead, we adopted the relevant parameters from  \citet{2014MNRAS.443.2391M},
who analysed a series of high quality transit light curves of Qatar-2\,b (see Table \ref{t:bestfitpar}).
Specifically, we used the stellar density $\rho_\star$=  2.240 $\pm$ 0.023 \gcm{ } from \citet{2014MNRAS.443.2391M},
together with our spectroscopic determinations of $T_{\rm eff}$ and [Fe/H], as input to 
the Yonsei-Yale evolutionary tracks to estimate the stellar mass $M_\star$=0.798 $\pm$ 0.040 M$_\odot$
and radius $R_\star$=0.793 $\pm$ 0.024 R$_\odot$. Then, we consistently scaled the values of the 
planet mass  $M_{\rm p}$=2.616 $\pm$ 0.071 M$_{\rm J}$ and radius $R_{\rm p}$=1.281 $\pm$ 0.039 R$_{\rm J}$.
As a check, we verified that the light curve model, calculated adopting those parameters, adjusts well
to the first half of our IAC80 photometric data (see top panel in Fig. \ref{Fig:fig_QATAR2_LC_RM_02}).

The RV time-series covering the Qatar-2\,b transit was fitted with the RM model, setting
free only three parameters: $\lambda$, $V\sin{I_\star}$ and $\gamma$. The best-fit values are 
reported in Table \ref{t:bestfitpar} and the RV measurements with the best-fit solution 
are displayed in the middle panel of Fig. \ref{Fig:fig_QATAR2_LC_RM_02}.
With only 8 in-transit data points and relatively large RV errors, as compared with the amplitude
of the RM effect, our detection may not appear statistically significant.
However, we note that the best-fit value of $V\sin{I_\star}$ = 2.09 $\pm$ 0.58 \kms\, agrees with 
the value of 2.0 $\pm$ 1.0 \kms\, derived from our spectroscopic analysis.
In order to assess the reliability of our estimate of $\lambda$=15$\pm$20\,deg, we then computed
RM models with $V\sin{I_\star}$ fixed to 2.0 \kms for three different values of $\lambda$ (90, 180, -90 deg),
shown in Fig. \ref{Fig:fig_QATAR2_LC_RM_02}. By comparing the different models, it appears that the data allow us to 
discriminate between different values of $\lambda$.
Furthermore, we note that the best-fit value of $\lambda$ we obtain is consistent with a previous determination 
of $\lambda$=4.3.$\pm$4.5 deg, obtained by \citet{2014MNRAS.443.2391M} using the spot crossing method.

\subsubsection{Stellar rotational period}\label{SubSec:DataAnalysis:Qatar2:Prot}

\begin{figure}
  \resizebox{\hsize}{!}{\includegraphics{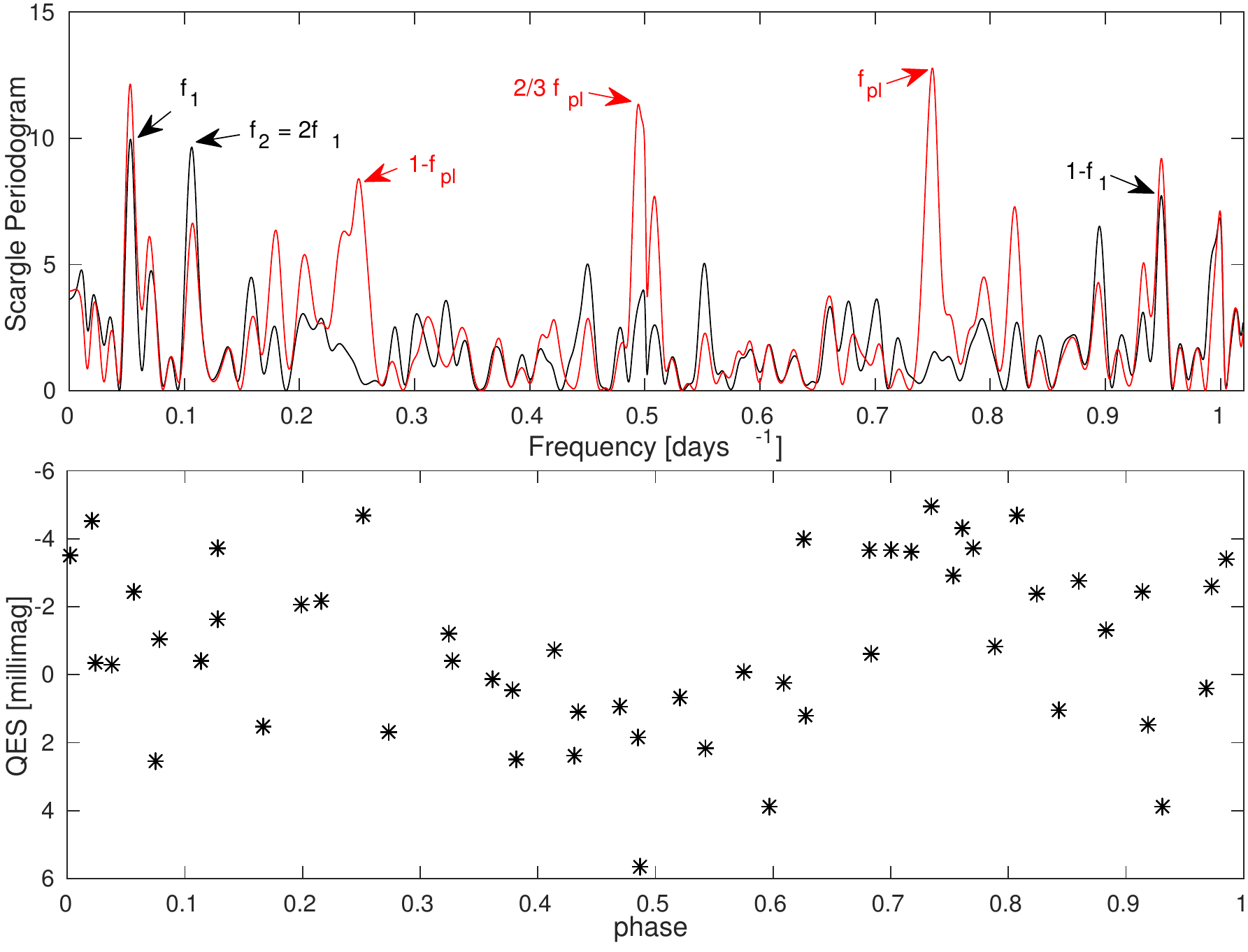}}
  \caption{Top panel: Scargle periodogram of the QES photometric monitoring data of Qatar-2, before (red line) and
  after (black line) removing the in-transit points. $f_{pl}$ corresponds to the planet orbital period; we deem the peak at $f_{1}$ to 
  originate from a photometric modulation at the stellar rotational period.  Bottom panel: Nightly averaged photometric data 
  phase-folded at the $P_{1}\equiv1/f_{1}$=18.77 days period.
  }
  \label{Fig:QATAR2_Prot}
\end{figure}

 We retrieved the publicly available QES (Qatar Exoplanet Survey) photometric monitoring data of Qatar-2 \citep{2012ApJ...750...84B}.
 The time-series consists of 1217 data points distributed over 54 nights and spanning an interval of 84 days.
 After removing some evident outliers (44 points),
 we analysed the remaining points for possible periodic signals, both before and after removing
 the in-transit data ($-0.04$<phase<0.04; 94 points).
The Lomb-Scargle periodograms (top panel in Fig. \ref{Fig:QATAR2_Prot}) show two peaks which become prominent after discarding
the in-transit points, corresponding to periods $P_1\sim$18.7 days and $P_2 \sim P_{1}/2$\,. \,
In order to assess the significance of these peaks, we computed the periodograms 
for 10$^5$ mock data sets, obtained by randomly permuting the time-stamps of the original data set.
Then, we calculated false alarm probability (FAP) levels; for instance,
a 1$\%$ FAP level means that in 10$^3$ out of 10$^5$ cases a peak higher than that level
was found in the periodograms, over the entire frequency range [10$^{-3}$, 1.02] days$^{-1}$.
In this way, we estimate the FAP associated with the peak at $P_1$ to be 0.9$\%$.
In the bottom panel of Fig. \ref{Fig:QATAR2_Prot} we show the Qatar-2 QES photometry, binned on a night-by-night basis,
phase-folded with the period $P_1$ and an arbitrarily chosen reference epoch.
We consider $P_1$ as the stellar rotational period. The light curve shows a clear minimum
around phase 0.5  and then a rather flat maximum from 0.8 to 0.3. The strong harmonic
 at $P_{1}/2$ observed in the periodogram also points to an asymmetric light curve.
 This shape can be explained by  a single cold spotted region visible for half period
  only, perfectly compatible with the equator-on orientation of our line of sight.
We used a bootstrap method, applied to the binned data, to derive the uncertainty on the period,
 obtaining $P_{\rm rot}$ = 18.77 $\pm$ 0.29 days.
 Very recently two independent works \citep{2016arXiv160807524M,2016arXiv160901314D} reported on the analysis of 
 K2 photometric time-series for Qatar-2, and both determined a stellar rotational period in agreement with our result,
 and confirmed that the planet is aligned.
 
 Using the values of $P_{\rm rot}$, $V\sin{I_\star}$, and $R_\star$, we derive that 
$I_\star$>58 deg with a 68 \% confidence level.
 From our determination of $\lambda$=15$\pm$20 deg, we estimate that the true
 spin-orbit angle is  $\varPsi$<43 deg with a 68 \% confidence level.


%

\section{Discussion} \label{Sec:Disc}

\subsection{Stellar activity}
It has been speculated that the presence of close in giant planets 
could cause an enhancement of the activity level of their host stars,
via star-planet tidal interactions and/or magnetic coupling 
\citep{2000ApJ...533L.151C,2008A&A...487.1163L,2012A&A...544A..23L}.
Many observational studies have looked for correlations between
stellar activity indicators and the presence and properties of exoplanets (e.g.  \cite{2015ApJ...811L...2M,2015A&A...578A..64B}).
\citet{2013ApJ...766....9S}, considering the 272 known FGK planetary hosts observed by GALEX,
found only tentative evidence that hot-Jupiters host stars are more FUV-active; \citet{2010A&A...515A..98P},
analyzing planet-bearing stars within 30 pc, concluded that there are
no correlations of X-ray luminosity or the activity indicators $L_{\rm X}/L_{\rm bol}$ with planetary
parameters.
\citet{2012A&A...540A..82K} found
statistically significant evidence that the equivalent width of the Ca II K line emission and log($R'_{\rm HK}$) activity 
parameter of the host star vary with the mass and orbital semi-major axis of the planet.
In a similar study, which considered however a sample of planets at larger orbital separations,
\citet{2011A&A...530A..73C} had not found significant correlations.

We have found that HAT-P-20 and WASP-43 have values of log($R'_{\rm HK}$) (see Table\,\ref{t:bestfitpar})
that place them among the most active planet-host stars, reinforcing the \citet{2012A&A...540A..82K} results
(see in particular their Figure 3 and 5).
\citet{2015MNRAS.452.2745S}, by studying a sample of stars monitored with the HARPS spectrograph for which
the presence of hot-Jupiters can be excluded with high confidence, found an empirical tight correlation
between the stellar rotational period $P_{\rm rot}$ and the average log($R'_{\rm HK}$)
 (see also \cite{1984ApJ...279..763N} and \cite{2008ApJ...687.1264M}).
According to their Equation (9), based on the measured $P_{\rm rot}$, WASP-43 and HAT-P-20
should have a log($R'_{\rm HK}$) of $-$4.62$\pm$0.07 and $-$4.57$\pm$0.07 respectively, lower than
the measured values  by 0.27$\pm$0.12 and  0.17$\pm$0.08 respectively.
Thus, we have found that, at the 2-$\sigma$ level of confidence,  both WASP-43 and HAT-P-20 
show an enhanced level of chromospheric activity as measured by the log($R'_{\rm HK}$) index.
The excess of activity could be an effect of the young age of the stars. Due to the late spectral type of the stars, 
 evolutionary models are unable to provide stringent constraints on their age. However, we calculated the
 Galactic space velocities (see Table \,\ref{t:bestfitpar}) using the spectroscopic parallaxes reported in 
 \citet{2011A&A...535L...7H} and  \citet{2011ApJ...742..116B} for WASP-43 and HAT-P-20, respectively.
 Both stars have space velocities not compatible with any of the known nearby young moving groups \citep{2004ARA&A..42..685Z}.
We conclude that the origin of the enhanced activity is likely to be sought in the tidal and/or magnetic
interactions of the stars with their close-in massive planetary companions.

\subsection{Obliquity and stellar rotation}
\begin{figure*}
  \resizebox{\hsize}{!}{\includegraphics{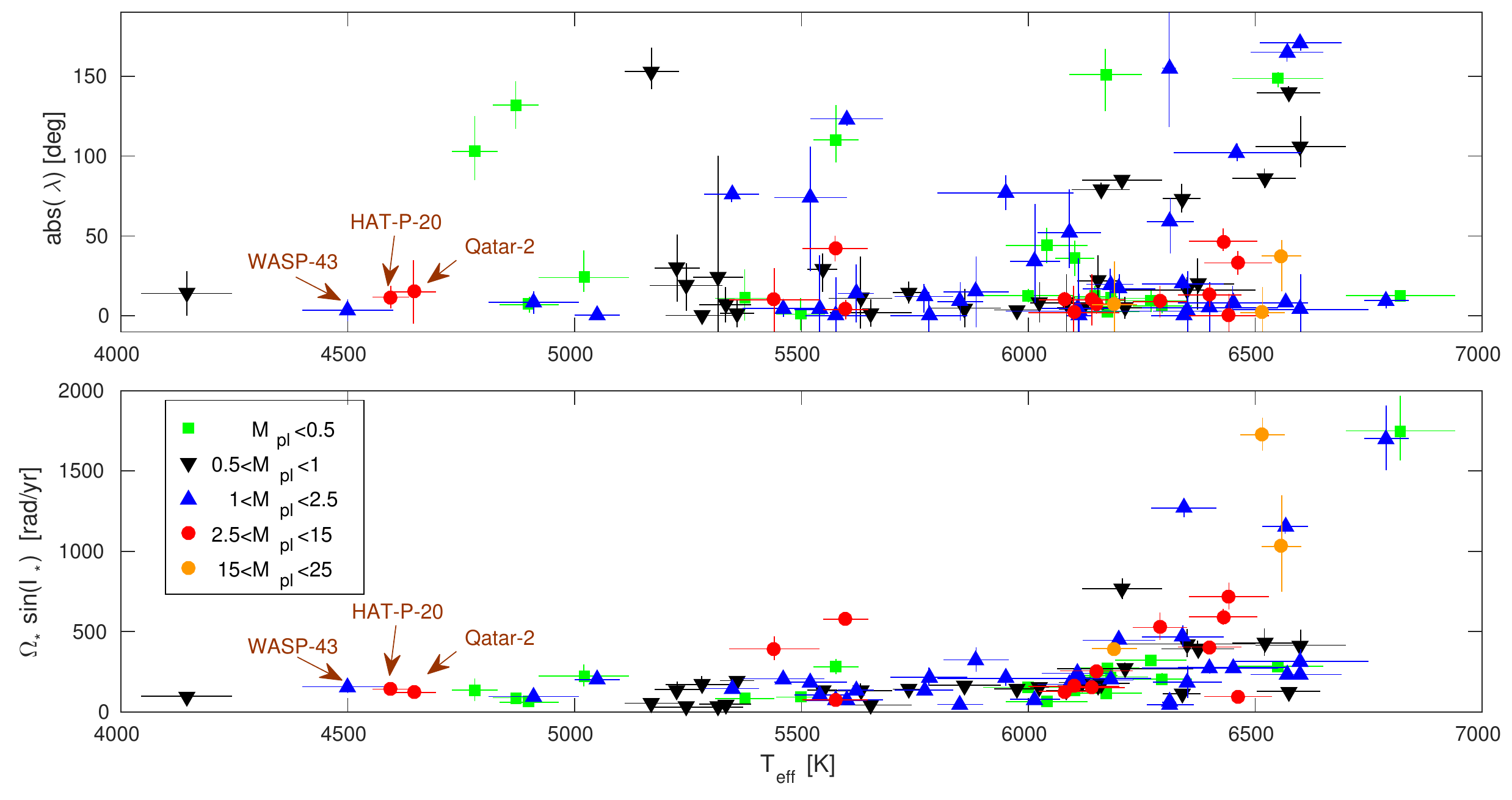}}
  \caption{Top panel: $\lambda$-$T_{\rm eff}$ diagram for all the transiting planets with published determinations of $\lambda$. Only two systems
  with $T_{\rm eff}$>7000\,K were not included. Bottom panel: Projected stellar angular rotational velocity $\Omega_* \sin(I_*)$ as a function of
  the stellar $T_{\rm eff}$. We referred to the http://www.astro.keele.ac.uk/jkt/tepcat/ website for
  the values of $\lambda$, $T_{\rm eff}$, $R_{\rm *}$ and $M_{\rm pl}$; the values of $V\sin{I_\star}$ were compiled by exploring the literature.
  }
  \label{Fig:Dawson_updated}
\end{figure*}

In Fig. \ref{Fig:Dawson_updated} we have revisited and updated the diagrams in Fig. 1 of 
\citet{2014ApJ...790L..31D}. Our diagrams also differ in that we did not apply 
the cuts in planet mass ($M_{\rm pl}>0.5 M_{\rm J}$) and period ($P$>7 days) used in
\citet{2014ApJ...790L..31D}. We stress that all the three stars studied
in this work have $T_{\rm eff}\sim$\,4500 K, hence they populate a region of the 
$\lambda$-$T_{\rm eff}$ diagram that was largely unexplored before.
Our three targets show small obliquities, therefore, considering also the
relatively high mass and short orbital period of their planets, they appear consistent with the $\lambda$-$T_{\rm eff}$ trend, 
according to which planets around stars with $T_{\rm eff}\lesssim$\,6250 K have aligned orbits \citep{2012ApJ...757...18A}.

\citet{2014ApJ...790L..31D} noticed that for $T_{\rm eff}$\,$\lesssim$\,6000 K, out of 19 systems with known $\lambda$,
the 2 with the highest values of the projected stellar angular rotational velocity $\Omega_* \sin{I_*}$ corresponded to the most massive planets (CoRoT-2b and CoRot-18b)
\footnote{The third massive planet shown in Fig. \ref{Fig:Dawson_updated} is HD80606b: with an orbital period of $\sim$111 days,
it is not expected to have any significant tidal interactions with its host star.}.
She argued that this was the result of a planet mass dependence of the stellar projected rotation frequencies for cool stars,
with massive planets being able to spin up their host stars via tidal interaction, contrasting the magnetic braking.
She was able to reproduce this and other observational trends in a theoretical framework in which, for cool stars, the 
mass dependence was insensitive to the stellar $T_{\rm eff}$ (see right bottom diagram of her Fig. 1).
However, our targets, in particular HAT-P-20 and Qatar-2, in spite of the large mass of the planets, 
have small values of $\Omega_* \sin{I_*}$ (see the bottom panel of Fig. \ref{Fig:Dawson_updated}). 
This suggests that probably, at the temperatures of our targets ($T_{\rm eff}\sim$4500 K),
the magnetic braking dominates over tidal spin up; a larger number of similar systems should be studied to 
settle this issue.

\subsection{Tidal timescales}
All our systems are currently not synchronized, i.e., far from tidal equilibrium. 
In principle, only HAT-P-20 could reach a stable equilibrium because its total angular momentum 
is very close to the minimum value required according to \citet{1980A&A....92..167H},
while for WASP-43 and Qatar-2 it is smaller by a factor of about two. 
However, given the steady angular momentum loss produced by the  magnetized wind in late-type stars, 
such a stable equilibrium, even if  established, cannot be maintained in HAT-P-20 \citep{2015A&A...574A..39D}. 
Since the orbital periods of the three systems are shorter than the rotation periods, tides transfer angular momentum from the orbit 
to the stellar spin and the fate of these close-in planets is to fall towards their host stars.

An estimate of the infall timescale in our systems is made very uncertain by our ignorance of the physics of the dissipation 
of  tidal kinetic energy inside late-type stars and planets, in particular of the dissipation of dynamical tides, 
i.e., the wave-like perturbations excited by the tidal potential that varies periodically in the reference frame 
of the stars and planets \citep{2014ARA&A..52..171O}. We parameterize the efficiency of tidal dissipation by means 
of the so-called modified tidal quality factors, $Q^{\prime}_{\rm s}$ and $Q^{\prime}_{\rm p}$ for the star and the planet, respectively.
In close binary systems consisting of two late-type stars, the observations suggest $Q^{\prime}_{\rm s} \approx 10^{6}$ \citep{2007ApJ...661.1180O} 
that would imply a short remaining lifetime for all our three systems, ranging from 
$\sim 13$~Myr for WASP-43 to $\sim 900$~Myr for HAT-P-20 \citep[from Eq.~(1) in]{2012MNRAS.425.2778M}. 
The e-folding timescale for the decrease of any initial obliquity of our systems would be comparable or shorter, 
ranging from $\sim 10$~Myr for WASP-43 to 160 Myr for HAT-P-20, if we adopt the constant time-lag model of \citet{2010A&A...516A..64L} 
and the presently measured parameters of the systems (cf. Table \ref{t:bestfitpar})\footnote{We computed the 
constant time lag $\tau$ using Eqs.~(18) and~(19) in \citet{2010A&A...516A..64L} and assuming that the planet rotation is synchronous with the orbit.}. 

An observational lower limit  $Q^{\prime}_{\rm s} \ga 10^{5}$ for WASP-43 has been obtained by \citet{2016AJ....151..137H}, 
but it is far too low to be useful in our context. Assuming a likely tidal spin-up of the host star in the HATS-18 system, 
\citet{2016arXiv160600848P} estimate $Q^{\prime}_{\rm s}$ for that solar-like host. Scaling their value according to the different 
rotation periods of our target host stars based on the inertial wave dissipation model by \citet{2007ApJ...661.1180O}, 
we assume $Q^{\prime}_{\rm s} = 10^{7}$, while $Q^{\prime}_{\rm p} = 10^{6}$ is appropriate for WASP-43\,b and Qatar-2\,b, 
and $Q^{\prime}_{\rm p} = 5 \times10^{6}$ for HAT-P-20\,b in view of its longer period, 
by scaling from the  value of $Q^{\prime}_{\rm p} \simeq 10^{5}$ for Jupiter \citep[see, Sect.~5.4 in]{2014ARA&A..52..171O}. 
A $Q^{\prime}_{\rm s} \sim 10^{7}$ is also in general agreement with the statistical study by \citet{2009ApJ...698.1357J}. 
With such values of the modified tidal quality factors, we obtain remaining lifetimes ranging 
from $\sim 130$~Myr for WASP-43 to $\sim 9$~Gyr for HAT-P-20. The expected $O-C$ of the epoch of mid-transit 
with respect to a constant-period ephemeris is the largest for WASP-43 and it is of $-5.5$~s in ten years. The e-folding decay 
time of the initial obliquity ranges from $\sim 100$~Myr for WASP-43 to $\sim 1.6$~Gyr for HAT-P-20, i.e., shorter 
than the probable ages of the systems, thus suggesting that tides could have had enough time to align any initially 
oblique spin of the stars. On the other hand, the dissipation of any initial orbital eccentricity is dominated by  
the tides inside the planets and the corresponding e-folding timescales are shorter than $\sim$10~Myr for WASP-43 and Qatar-2, while 
it is of $\sim$4~Gyr in the case of HAT-P-20, suggesting that its present non-zero eccentricity could have been excited when the system formed. 

As mentioned in Sec. \ref{SubSec:PhRed}, HAT-P-20 has a visual companion at an angular separation of 6.9 arcsec.
In the URAT catalog \citep{2015yCat.1330....0N,2015AJ....150..101Z}, HAT-P-20 and its companion are reported to have a common proper motion,
separation $\rho$\,=\,6.93 arcsec and position angle PA\,=\,320.6 deg at the epoch TE\,=\,2013.865 yr.
Previous observations indicated similar values:\,$\rho$\,=\,6.939 arcsec, PA\,=\,320.6 deg,  TE\,=\,2001.068 yr \citep{2013AJ....146...76H};
\,$\rho$\,=\,6.860 arcsec, PA\,=\,320.3 deg,  TE\,=\,1998.07 yr \citep{2006AJ....132...50W}. Thus, in all probability
the two stars are physical companions; at an estimated distance of 70$\pm$3 pc \citep{2011ApJ...742..116B}, the two stars have a projected
separation of $\sim$490 AU. The stellar companion, based on its 2-MASS colors, is an M-dwarf \citep{2015A&A...576A..42S}.
It can not be excluded that gravitational perturbations from the companion played a role in the orbital evolution of HAT-P-20\,b
leading to the current slightly eccentric and misaligned orbit.

%

\section{Conclusions} \label{Sec:Concl}

We have measured the RM effect for the close-in and massive transiting planets orbiting WASP-43, HAT-P-20 and Qatar-2,
three K-dwarf stars with effective temperatures of 4500, 4595 and 4640 K, respectively.
We have found that the true spin-orbit angle $\varPsi$ of WASP-43\,b is consistent with zero. 
For Qatar-2\,b, we only report a marginal detection of the RM effect, however our results support previous evidence
that the system is aligned.
HAT-P-20\,b, the one among the three with the largest semi-major axis, 
has small but significant eccentricity ($e$\,=\,0.0172 $\pm$0.0016) and obliquity ($\varPsi$\,=\,36 $_{-12}^{+10}$ deg)
which might be related to the presence of the stellar companion at projected separation of $\sim$490 AU.

HAT-P-20 and WASP-43, for which we could obtain  reliable measurements of the average $R'_{\rm HK}$
index, show an activity level exceeding the values typical for stars with the same
rotation period. We take this as a likely manifestation of either tidal or magnetic star-planet interactions.
A larger number of spectra with higher S/N would be needed to study the $R'_{\rm HK}$ variability
and possible modulations with the planet orbital phase.

Contrarily to what has been observed in two stars with  $T_{\rm eff}\sim$5500 K hosting massive planets,
our targets do not show clear evidence of stellar rotational spin-up.

Overall, our findings are consistent with the scenario in which star-planet interactions have been
effective in circularizing and aligning the planetary orbits, similarly to what already observed for 
hotter stars up to $T_{\rm eff}\sim$6250 K.
%

\begin{acknowledgements} 
 We thank the anonymous referee for her/his useful comments and suggestions.
The GAPS project in Italy acknowledges the support by INAF through 
the "Progetti Premiali" funding scheme of the Italian Ministry of Education, University, and Research.
Based on observations collected at Copernico telescope (Asiago, Italy) of the INAF - Osservatorio Astronomico di Padova.
V. N. acknowledges partial support by the Universit\`{a} di Padova through 
the "Studio preparatorio per il Plato Input Catalog" grant (\#2877-4/12/15) funded by the ASI-INAF agreement (n.~2015-019-R.0).
The authors acknowledge Dr J.M. Alcal\'{a},
Dr. L. Bedin and Dr. G. Lodato for their comments and suggestions.
\end{acknowledgements}
%

\bibliographystyle{aa} 
\bibliography{ref_paper__GAPS_XIII.bib} 
%

%
\begin{appendix}

\section{RV tables}


\begin{table*}[th]
\caption{
 HARPS-N RV data for WASP-43. The columns report: BJD (TDB), the mid-exposure Barycentric Julian Dates in Barycentric Dynamical Time; 
T$_{\rm exp}$, the exposure time; RV and error are the radial velocity measurement and its estimated uncertainty; FWHM, the Full Width at
Half Maximum of the Cross-Correlation Function; Bis. Span, the radial velocity bisector span of the CCF; Airmass, the airmass of the star at the beginning of the exposure;
Flag, indicating wether the spectrum was taken in-transit (i) or off-transit (o).
}
\label{t:WASP43-RV-data}
\centering
\begin{tabular}{rrrrrrrc}
\hline
\hline
 BJD (TDB)      & T$_{\rm exp}$   &  RV      & error  &    FWHM   & Bis. Span &  Airmass & Flag \\ 
                & [sec]       &[\kms]    & [\kms] &    [\kms] & [\kms]    &          &                      \\

\hline
 2456363.415882  &     450  & 	$-$3.2736  &  0.0111  &  7.17  &    0.086  &  1.59  &  o \\    
 2456363.421396  &     450  & 	$-$3.3145  &  0.0120  &  7.15  &    0.076  &  1.55  &  o \\	 
 2456363.426923  &     450  & 	$-$3.3185  &  0.0115  &  7.16  &    0.027  &  1.51  &  o \\	 
 2456363.432437  &     450  & 	$-$3.3278  &  0.0096  &  7.18  &    0.015  &  1.48  &  o \\	 
 2456363.437951  &     450  & 	$-$3.3683  &  0.0093  &  7.14  &    0.060  &  1.45  &  o \\	    
 2456363.443460  &     450  & 	$-$3.3820  &  0.0091  &  7.16  &    0.056  &  1.42  &  o \\	   
 2456363.448974  &     450  & 	$-$3.4072  &  0.0092  &  7.25  &    0.046  &  1.40  &  o \\	    
 2456363.454484  &     450  & 	$-$3.4373  &  0.0093  &  7.16  &    0.056  &  1.38  &  o \\	 
 2456363.459988  &     450  & 	$-$3.4560  &  0.0088  &  7.17  &    0.057  &  1.36  &  o \\	 
 2456363.465502  &     450  & 	$-$3.4526  &  0.0095  &  7.16  &    0.102  &  1.34  &  i \\    
 2456363.471016  &     450  & 	$-$3.5011  &  0.0099  &  7.12  &    0.034  &  1.33  &  i \\	 
 2456363.476525  &     450  & 	$-$3.4876  &  0.0096  &  7.16  &    0.079  &  1.31  &  i \\   
 2456363.482039  &     450  & 	$-$3.5192  &  0.0096  &  7.20  &    0.057  &  1.30  &  i \\	 
 2456363.487549  &     450  & 	$-$3.5517  &  0.0094  &  7.19  &    0.057  &  1.29  &  i \\	   
 2456363.493054  &     450  & 	$-$3.5978  &  0.0093  &  7.17  &    0.080  &  1.29  &  i \\	    
 2456363.498567  &     450  & 	$-$3.6221  &  0.0102  &  7.14  &    0.024  &  1.28  &  i \\	   
 2456363.504081  &     450  & 	$-$3.6596  &  0.0091  &  7.15  &    0.071  &  1.28  &  i \\	    
 2456363.509590  &     450  & 	$-$3.6773  &  0.0086  &  7.14  &    0.023  &  1.28  &  i \\	    
 2456363.515104  &     450  & 	$-$3.6747  &  0.0090  &  7.19  &    0.060  &  1.28  &  i \\	    
 2456363.520614  &     450  & 	$-$3.6976  &  0.0097  &  7.23  &    0.005  &  1.28  &  o \\	 
 2456363.526123  &     450  & 	$-$3.7218  &  0.0103  &  7.18  &    0.032  &  1.28  &  o \\    
 2456363.531624  &     450  & 	$-$3.7571  &  0.0109  &  7.22  &    0.025  &  1.29  &  o \\   
 2456363.537128  &     450  & 	$-$3.7764  &  0.0113  &  7.17  &    0.043  &  1.30  &  o \\	 
 2456363.542642  &     450  & 	$-$3.7869  &  0.0104  &  7.13  &    0.062  &  1.30  &  o \\	    
 2456363.548156  &     450  & 	$-$3.8216  &  0.0113  &  7.21  &    0.050  &  1.31  &  o \\	    
 2456363.553670  &     450  & 	$-$3.8339  &  0.0132  &  7.18  &    0.076  &  1.33  &  o \\    
 2456363.559179  &     450  & 	$-$3.8659  &  0.0117  &  7.16  &    0.073  &  1.34  &  o \\	   
 2456363.564693  &     450  & 	$-$3.8770  &  0.0154  &  7.22  &    0.106  &  1.36  &  o \\	
 2456363.571207  &     450  & 	$-$3.8913  &  0.0144  &  7.26  &    0.083  &  1.38  &  o \\	 
 2456363.576712  &     450  & 	$-$3.9012  &  0.0121  &  7.18  &    0.068  &  1.40  &  o \\	    
 2456363.582230  &     450  & 	$-$3.9561  &  0.0163  &  7.17  &    0.082  &  1.43  &  o \\	 
 2456363.587740  &     450  & 	$-$3.9589  &  0.0127  &  7.15  &    0.048  &  1.46  &  o \\	    
 2456376.517002  &     900  & 	$-$3.6434  &  0.0058  &  7.27  &    0.034  &  1.33  &  i \\	   
 2456377.589657  &     900  & 	$-$4.1247  &  0.0088  &  7.40  & $-$0.093  &  1.81  &  o \\	 
 2456381.558437  &     450  & 	$-$4.0933  &  0.0088  &  7.19  &    0.072  &  1.59  &  o \\	 
 2456381.563762  &     450  & 	$-$4.1130  &  0.0089  &  7.18  &    0.030  &  1.64  &  o \\	 
 2456382.512779  &     900  & 	$-$3.9513  &  0.0052  &  7.20  &    0.041  &  1.37  &  o \\   
 2457050.557137  &     900  & 	$-$3.2261  &  0.0192  &  7.18  &    0.101  &  1.43  &  o \\	
 2457116.447829  &     900  & 	$-$3.2458  &  0.0045  &  7.11  &    0.075  &  1.28  &  o \\	 
 2457145.424807  &     900  & 	$-$4.1267  &  0.0154  &  7.02  &    0.052  &  1.38  &  o \\    
\hline 
\end{tabular}
\end{table*}



\begin{table*}
\caption{
HARPS-N RV data for HAT-P-20. See the caption of Table \ref{t:WASP43-RV-data} for the meaning of the columns.
}
\label{t:HATP20-RV-data}
\centering
\begin{tabular}{rrrrrrrc}
\hline
\hline
 BJD (TDB)      & T$_{\rm exp}$   &  RV      & error  &    FWHM   & Bis. Span &  Airmass & \tablefootmark{\dag} \\ 
                & [sec]       &[\kms]    & [\kms] &    [\kms] & [\kms]    &          &                      \\
\hline

2456252.743420  &   900  &  $-$17.8116    &  0.0034   &  7.02  &    0.073  &  1.05   & o \\
2456305.646766  &   900  &  $-$17.6177    &  0.0030   &  7.01  &    0.039  &  1.17   & o \\
2456323.675026  &   900  &  $-$19.2777    &  0.0082   &  7.12  &    0.046  &  1.79   & o \\
2456569.727872  &   900  &  $-$16.8530    &  0.0031   &  7.03  &    0.055  &  1.15   & o \\
2456581.752206  &   900  &  $-$17.8062    &  0.0026   &  7.05  &    0.050  &  1.02   & o \\
2456606.758475  &  1200  &  $-$16.9772	  &  0.0039   &  7.08  &    0.026  &  1.03   & o \\
2456608.768700  &   900  &  $-$19.0709    &  0.0025   &  7.10  &    0.064  &  1.05   & o \\
2456608.778237  &   400  &  $-$19.0600    &  0.0045   &  7.10  &    0.057  &  1.06   & o \\
2456609.779656  &   900  &  $-$16.8533	  &  0.0056   &  7.06  &    0.050  &  1.08   & o \\
2456621.746725  &  1200  &  $-$17.1818    &  0.0030   &  7.12  &    0.036  &  1.08   & o \\
2456622.723432  &   900  &  $-$19.3090    &  0.0035   &  7.12  &    0.056  &  1.03   & o \\
2456631.627655  &   900  &  $-$19.2637    &  0.0024   &  7.01  &    0.068  &  1.01   & o \\
2456719.481675  &   900  &  $-$17.1670    &  0.0063   &  7.11  &    0.054  &  1.08   & o \\
2456723.544801  &   900  &  $-$19.2887    &  0.0036   &  7.10  &    0.048  &  1.43   & o \\
2456728.410880  &   600  &  $-$17.8740    &  0.0046   &  7.05  &    0.009  &  1.01   & o \\
2456728.418111  &   600  &  $-$17.8955    &  0.0046   &  7.08  &    0.027  &  1.01   & o \\
2456728.425345  &   600  &  $-$17.9150    &  0.0046   &  7.07  &    0.017  &  1.02   & o \\
2456728.432580  &   600  &  $-$17.9264    &  0.0046   &  7.09  &    0.013  &  1.03   & o \\
2456728.439814  &   600  &  $-$17.9500    &  0.0050   &  7.09  & $-$0.014  &  1.04   & o \\
2456728.447035  &   600  &  $-$17.9705    &  0.0050   &  7.08  & $-$0.001  &  1.06   & i \\
2456728.454265  &   600  &  $-$17.9766    &  0.0046   &  7.09  &    0.016  &  1.07   & i \\
2456728.461495  &   600  &  $-$17.9831    &  0.0044   &  7.06  &    0.005  &  1.09   & i \\
2456728.468729  &   600  &  $-$18.0035    &  0.0041   &  7.07  &    0.033  &  1.11   & i \\
2456728.475959  &   600  &  $-$18.0383    &  0.0046   &  7.07  &    0.023  &  1.13   & i \\
2456728.483184  &   600  &  $-$18.0725	  &  0.0048   &  7.08  &    0.030  &  1.16   & i \\
2456728.490418  &   600  &  $-$18.0874    &  0.0057   &  7.07  &    0.030  &  1.18   & i \\
2456728.497653  &   600  &  $-$18.1251    &  0.0062   &  7.12  &    0.011  &  1.22   & i \\
2456728.504882  &   600  &  $-$18.1492    &  0.0052   &  7.07  &    0.027  &  1.25   & i \\
2456728.512125  &   600  &  $-$18.1571	  &  0.0052   &  7.06  &    0.015  &  1.29   & i \\
2456728.519346  &   600  &  $-$18.1668    &  0.0054   &  7.08  &    0.010  &  1.33   & i \\
2456728.526585  &   600  &  $-$18.1851    &  0.0056   &  7.05  &    0.009  &  1.38   & o \\
2456728.533816  &   600  &  $-$18.2109    &  0.0054   &  7.06  &    0.033  &  1.44   & o \\
2456728.541041  &   600  &  $-$18.2263    &  0.0065   &  7.16  & $-$0.006  &  1.50   & o \\
2456728.548267  &   600  &  $-$18.2480    &  0.0076   &  7.15  &    0.016  &  1.57   & o \\
2456728.555489  &   600  &  $-$18.2684    &  0.0071   &  7.17  & $-$0.008  &  1.65   & o \\
2456728.562718  &   600  &  $-$18.2802    &  0.0079   &  7.20  &    0.005  &  1.75   & o \\
2456728.569953  &   600  &  $-$18.3064    &  0.0091   &  7.16  & $-$0.018  &  1.85   & o \\
2456976.687433  &   900  &  $-$19.2139    &  0.0022   &  6.95  &    0.043  &  1.01   & o \\
2457005.717112  &   900  &  $-$18.7385    &  0.0043   &  6.94  &    0.072  &  1.13   & o \\
2457011.702769  &   900  &  $-$18.1656    &  0.0036   &  6.96  &    0.065  &  1.14   & o \\
2457125.412095  &   900  &  $-$18.4303    &  0.0020   &  7.04  &    0.068  &  1.25   & o \\
2457389.617073  &   900  &  $-$17.5685    &  0.0045   &  7.05  &    0.063  &  1.02   & o \\
\hline
\end{tabular}
\tablefoot{ 
\tablefoottext{\dag}{i\,$\equiv$\,in-transit, o\,$\equiv$\,out-of-transit}
}
\end{table*}



\begin{table*}
\caption{
HARPS-N RV data for Qatar-2.  See the caption of Table \ref{t:WASP43-RV-data} for the meaning of the columns.
}
\label{t:Qatar2-RV-data}
\centering
\begin{tabular}{rrrrrrrc}

\hline
\hline
 BJD (TDB)      & T$_{\rm exp}$   &  RV      & error  &    FWHM   & Bis. Span &  Airmass & \tablefootmark{\dag} \\ 
                & [sec]       &[\kms]    & [\kms] &    [\kms] & [\kms]    &          &                      \\

\hline
 2456775.434193  &   900  & $-$23.7480  &  0.0271  &  6.69  &	  0.089  &  1.51  &   o  \\	       
 2456775.444895  &   900  & $-$23.7285  &  0.0265  &  6.65  &	  0.025  &  1.44  &   o  \\	       
 2456775.455598  &   900  & $-$23.8385  &  0.0307  &  6.69  &	  0.081  &  1.38  &   o  \\	       
 2456775.466305  &   900  & $-$23.8248  &  0.0261  &  6.72  &  $-$0.004  &  1.34  &   o  \\	       
 2456775.477012  &   900  & $-$23.8530  &  0.0253  &  6.75  &	  0.089  &  1.30  &   o  \\	       
 2456775.487705  &   900  & $-$23.8691  &  0.0305  &  6.73  &  $-$0.069  &  1.27  &   i  \\	       
 2456775.498412  &   900  & $-$23.8725  &  0.0271  &  6.72  &	  0.132  &  1.25  &   i  \\	     
 2456775.509114  &   900  & $-$23.9181  &  0.0242  &  6.73  &  $-$0.011  &  1.24  &   i  \\	       
 2456775.519825  &   900  & $-$23.9703  &  0.0254  &  6.60  &	  0.038  &  1.23  &   i  \\	       
 2456775.530532  &   900  & $-$24.0137  &  0.0225  &  6.74  &  $-$0.008  &  1.23  &   i  \\	       
 2456775.541239  &   900  & $-$24.0664  &  0.0304  &  6.74  &	  0.012  &  1.23  &   i  \\	       
 2456775.551946  &   900  & $-$24.0915  &  0.0374  &  6.59  &	  0.111  &  1.24  &   i  \\	       
 2456775.562648  &   900  & $-$24.0793  &  0.0407  &  6.67  &	  0.055  &  1.26  &   i  \\	       
 2456775.573355  &   900  & $-$24.1343  &  0.0432  &  6.71  &	  0.015  &  1.28  &   o  \\	       
 2456775.584062  &   900  & $-$24.1286  &  0.0409  &  6.80  &	  0.000  &  1.32  &   o  \\	       
 2456775.594760  &   900  & $-$24.1363  &  0.0361  &  6.72  &  $-$0.030  &  1.35  &   o  \\	      
 2456775.605458  &   900  & $-$24.2294  &  0.0386  &  6.69  &	  0.052  &  1.40  &   o  \\	       

\hline 
\end{tabular}
\tablefoot{ 
\tablefoottext{\dag}{i\,$\equiv$\,in-transit, o\,$\equiv$\,out-of-transit}
}
\end{table*}


\end{appendix}
%

\end{document}